\documentclass[twocolumn]{aastex62}
\def\mod/{\textit{mod}$_{\text{struct}}$}

\begin{document}

\title{Stellar Evolution in Real Time: Models Consistent with Direct Observation of Thermal Pulse in T Ursae Minoris}
\shorttitle{Thermal Pulse in T~UMi}

\author[0000-0002-8159-1599]{L\'{a}szl\'{o} Moln\'{a}r}
\email{molnar.laszlo@csfk.mta.hu}
\affiliation{Konkoly Observatory, MTA CSFK, Budapest, Konkoly Thege Mikl\'os \'ut 15-17, Hungary}
\affiliation{MTA CSFK Lend\"ulet Near-Field Cosmology Research Group, 1121, Budapest, Konkoly Thege Mikl\'os \'ut 15-17, Hungary}

\author[0000-0002-8717-127X]{Meridith Joyce}
\email{meridith.joyce@anu.edu.au}
\affiliation{Research School of Astronomy and Astrophysics, Australian National University, Canberra, ACT 2611, Australia}

\author[0000-0002-3234-1374]{L\'aszl\'o L.\ Kiss}
\affiliation{Konkoly Observatory, MTA CSFK, Budapest, Konkoly Thege Mikl\'os \'ut 15-17, Hungary}
\affiliation{Sydney Institute for Astronomy, School of Physics A29, University of Sydney, NSW 2006, Australia}

\shortauthors{Moln\'{a}r, Joyce \& Kiss}

\label{firstpage}

\begin{abstract}
Most aspects of stellar evolution proceed far too slowly to be directly observable in a single star on human timescales.  The thermally pulsing asymptotic giant branch is one exception. The combination of state-of-the-art modelling techniques with data assimilated from observations collected by amateur astronomers over many decades provide, for the first time, the opportunity to identify a star occupying precisely this evolutionary stage.
In this study, we show that the rapid pulsation period change and associated reduction in radius in the bright, northern variable star T~Ursae Minoris are caused by the recent onset of a thermal pulse. 
We demonstrate that T~UMi transitioned into a double-mode pulsation state, and we exploit its asteroseismic features to constrain its fundamental stellar parameters.
We use evolutionary models from MESA and linear pulsation models from GYRE to track simultaneously the structural and oscillatory evolution of models with varying mass. We apply a sophisticated iterative sampling scheme to achieve time resolution $\le10$ years at the onset of the relevant thermal pulses. 

We report initial mass of $2.0\pm0.15\, \mathrm{M}_\odot$ and an age of $1.17 \pm 0.21$ Gyr for T~UMi. This is the most precise mass and age determination for a single asymptotic giant branch star ever obtained. 
The ultimate test of our models will be the continued observation of its evolution in real time: we predict that the pulsation periods in T~UMi will continue shortening for a few decades before they rebound and begin to lengthen again, as the star expands in radius.
\end{abstract}


%%%%%%%%%%%%%%%%%%%%%%%%%%%%%%%%%%%%%%%%%%%%%%%%%%

%%%%%%%%%%%%%%%%% BODY OF PAPER %%%%%%%%%%%%%%%%%%

\section{Introduction}

The Asymptotic Giant Branch (AGB) is a short but important phase in the lives of low-- and intermediate--mass (${\sim}0.5\text{--}8.0\, \mathrm{M}_{\odot}$) stars \citep{Karakas2017}. 
AGB stars reach their peak luminosities and lose increasing amounts of mass to the interstellar medium during this period, before shedding their envelopes entirely and collapsing into white dwarfs. Most stars also undergo repeated helium shell flashes, or thermal pulses (TP), during the AGB. The thin, helium-burning shell around the inert C/O core is thermally unstable, and runaway burning episodes occur when enough helium has built up under the hydrogen-burning shell. The pulses may also dredge up fusion products to the surface, if the convection zone reaches the helium-burning shell during the pulse. This is known as the third dredge-up (3DU). The TP-AGB evolutionary phase extends to about 8~$\mathrm{M}_\odot$, although the transition and boundaries between AGB, super-AGB, and high-mass stars that evolve to core-collapse supernovae are still actively researched \citep{doherty2017}. 

\subsection{Pulsating AGB stars}

Many AGB stars are pulsating stars, too, and these are usually categorized as ``Mira'' or semiregular variables. We refer to the He-shell flash episodes as \textit{pulses} and to coherent global oscillations in the envelope as \textit{pulsations} throughout the paper in order to avoid conflating these phenomena. The distinction between these classes is largely phenomenological, based on their pulsation properties: Miras vary consistently in brightness with very large amplitudes that exceed 2.5~mag in the optical band, whereas semiregulars exhibit more irregular, lower amplitude variations. They are closely linked by evolution, and changes to the stellar structure, especially during a thermal pulse, may considerably change the pulsation properties and could shift a star between the two classes or temporarily quench the pulsation entirely \citep[see, e.g.,][]{kerschbaum-1992}. 

Given the extended envelopes and large sizes of AGB stars, they exhibit slow pulsations with periods measured in months to years. Their high intrinsic brightness and large-amplitude variations make them easy to follow, even with visual estimates. The long periods, however, mean that several decades of sustained observation are required in order to detect changes. Fortunately, amateur and professional observations now span more than a century for many of these stars, allowing us to investigate such changes in their pulsation.

Pulsation variations in Miras have long been known to manifest both as cycle-to-cycle changes and longer, secular variations. The former are generally attributed to temporal variations of large, hot cells in the photospheres of the stars that break the spherical symmetry of the pulsation and disrupt the extended atmospheres surrounding the star.
Changing hot spots were directly observed on the surface of $\chi$ Cyg, via interferometric measurements \citep{Lacour2009}.

Variations longer than the pulsation have also been observed in several Mira stars. Some of the stars with period changes are the so-called ``meandering" Miras, whose periods do not change monotonically but quasi-periodically, on timescales of 10--80 yrs. The origin of these variations is not well understood and may not be evolutionary at all \citep{templeton2005}. 
In a few cases, variations in the pulsation were attributed to the presence of chaos, driven by energy exchange feedback between pulsation modes \citep{buchler-2004}. In other cases, however, the apparent changes in pulsation period and amplitude---especially in semiregular stars,---are caused simply by the beating of two different pulsation modes \citep{kiss1999,kiss-2000}. We note that in some cases, these stars cross the Mira/semiregular amplitude limit repeatedly. 

Nevertheless, some Mira stars must inherently exist in the process of a thermal pulse. Multiple stars have been proposed to be undergoing this process, such as R~Aql, R~Hya and W~Dra. \citet{wood-zarro-1981} related the strong period changes seen in these three stars to the luminosity changes present in their early model calculations of the TP-AGB phase. However, it is not always straightforward to relate changes in the pulsation properties to thermal pulses. Dramatic changes in the pulsation of R~Dor, for example, are attributed to mode switching without any signs of a TP happening in the star \citep{rdor-1998}. 
In fact, it is easier to determine whether a star has recently undergone a TP---i.e., if the $s$--process elements are visible in the spectrum---than it is to deduce that a star is entering one. Changes or peculiarities in the abundance profile can signal an ongoing or recently finished TP if the 3DU has been activated \citep{lxcyg-2016,tumi-uttenthaler2011}.

The onset of a TP can be signaled by a rapid decrease in the pulsation period. At the start of a TP, the luminosity of the helium-burning shell spikes rapidly. This excess energy is used up almost entirely to expand the inner regions of the star. The expanding H-burning shell above cools so much that it is temporarily extinguished, which causes the outer layers of the star to shrink and the surface luminosity to drop at the beginning of the TP \citep{schwarzschild1967}. This initial phase may last for several decades. The decrease in radius, in turn, causes the pulsation frequencies to shorten, providing an efficient proxy to detect the TP. The most spectacular change among Miras has been observed in T Ursae Minoris (hereafter T~UMi). Currently, T~UMi is the best  candidate for a star at the onset of a pulse \citep{templeton2005}. 

\begin{figure*}
	\includegraphics[width=\textwidth]{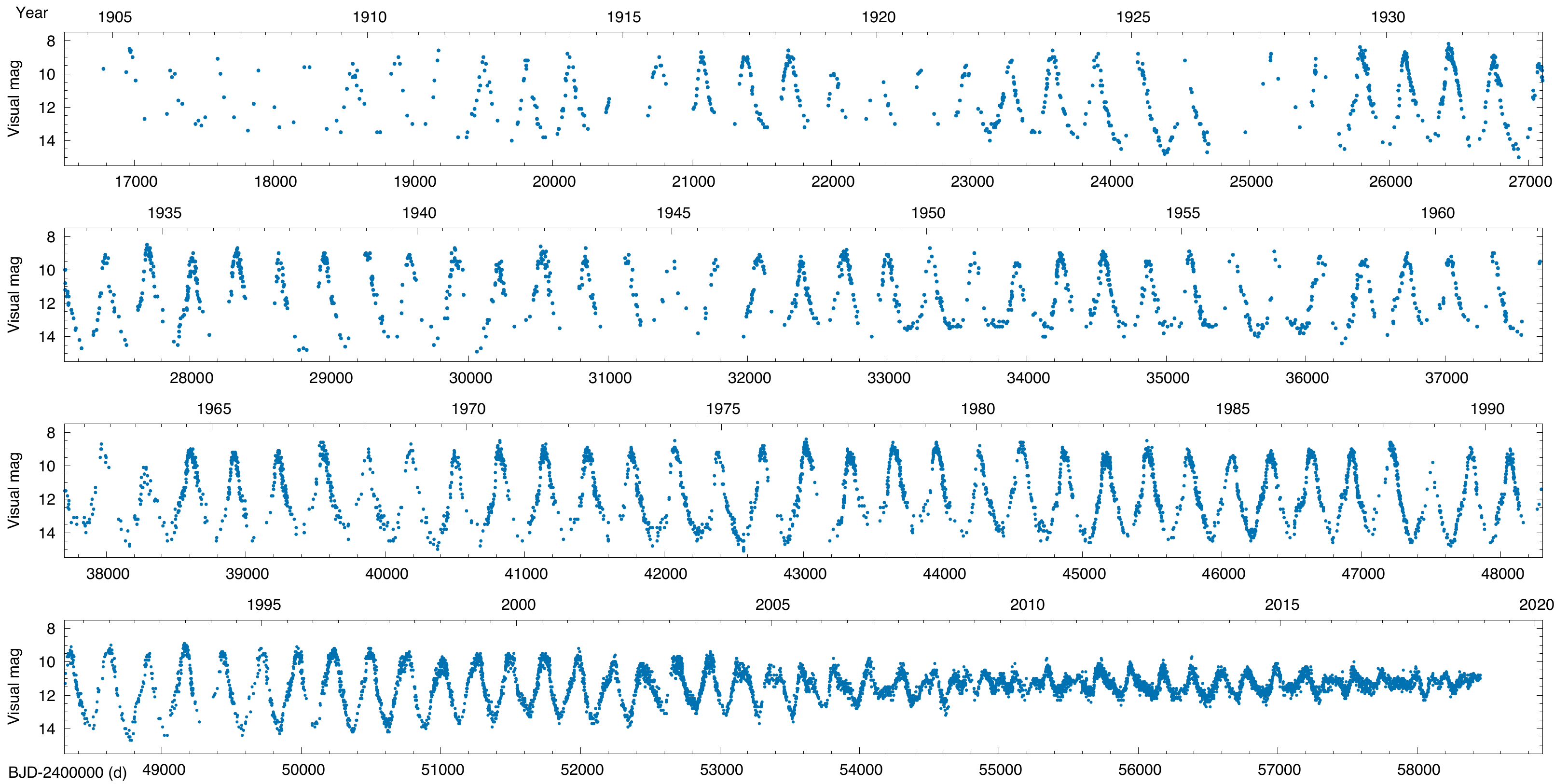}
    \caption{AAVSO visual light curve of T~UMi. Changes in the overall shape and amplitude of the pulsation are apparent in the lowest panel.}
    \label{lightcurve}
\end{figure*}

\subsection{T Ursae Minoris}

The light variations of T~UMi, a $V=11.5$ mag, M-type, single star, were first reported by \citet{Campbell-1912}. The star has been observed for more than a century, both by professional and amateur astronomers, mostly using photographic data and visual estimates. For much of this time, the Mira-type pulsations of T~UMi were unremarkable, displaying only some undulations between 300--330 d with an amplitude of 4.0--5.5 mag in the optical band. The onset of rapid period decrease was first noted by \citet{tumi-gal1995} and \citet{tumi-mattei1995}. Since then, the star has been more closely monitored, and CCD observations, mostly in the Johnson $V$ band, have also been collected over the last 20 years \citep[e.g.,][]{smelcer2002,smelcer2006}.

The possible reasons behind the steep decline of the pulsation period were first investigated in detail by \citet{tumi-szatmary2003}. They derived a rate of $-3.8 \pm 0.4$ d/y for the period change and found a slight decreasing trend in the pulsation-averaged intensity of the star. They concluded that the star is probably at the onset a TP, when the stellar radius and luminosity are declining, and that the change in radius is responsible for the sudden drop in the pulsation period. They also raised the possibility of the star simply switching from first-overtone to second-overtone pulsation. The latter, however, would have required the pulsation period to stabilize in the immediate future, and this conclusion is thus testable with continued observations. Interestingly, \citet{foster2010} later suggested in a short note that the star may be transitioning into a double-mode state that has been seen in multiple semiregular stars, but this finding was not detailed or further analyzed. Finally, \citet{tumi-uttenthaler2011} found that the periodicity of the star decreased to as short as 229~d, then seemed to jump suddenly to 113.6~d. This would then indicate a mode switch from the fundamental to the first-overtone mode, possibly triggered by the TP. 

Although comparisons between TP model calculations and stars existed, as, for example, done by \citet{wood-zarro-1981} or \citet{uttenthaler2016}, the onset of a pulse is not easy to handle. In order to model accurately the initial drop in radius and luminosity, the temporal resolution has to be on the decadal scale. As \citet{templeton2005} pointed out, models in the 1990s usually used time steps in the order of centuries---far too sparse for a quantitative comparison with the changes seen in T~UMi. The only attempt so far to model T~UMi both in terms of evolution and pulsation was done very recently by \citet{fadeyev2018}. He found that a $1.2\, \mathrm{M}_\odot$ initial-mass model fit the observations best, transitioning from a fundamental-mode pulsation around 315~d to first-overtone pulsation near 114~d. He also derived physical parameters, obtaining a luminosity of $4080\, \mathrm{L}_\odot$, a radius of $220\, \mathrm{R}_\odot$, and an age of $4.3 \times 10^9$~yr. However, this analysis did not assess the possibility of T~UMi experiencing a mode switch without an associated TP.

In this paper, we analyze the new observations gathered since the work of \citet{tumi-szatmary2003} in detail to determine whether the star has undergone a mode switch and whether the decline of the pulsation period has ended (Sect.~\ref{sec:obs}). We then apply the observational constraints to TP-AGB evolutionary and seismic models spanning a range of masses in order to provide a quantitative comparison more detailed than any previous modelling in Sect.~\ref{sec:calc}. We come to the conclusion that T~UMi is indeed at the onset of a thermal pulse, and provide testable, tightly constrained estimates of its fundamental parameters in Sect.~\ref{sec:results}. Finally, in Section \ref{sec:summary} we provide observationally verifiable predictions of its behavior over the next few decades as a function of our initial mass estimates. 

\section{Observations}
\label{sec:obs}
The onset of a TP in T~UMi was initially proposed largely based on the rapid period change the star has undergone. In addition to processing recent observations, we re-analyzed the entire visual light curve currently available at the database of the American Association of Variable Star Observers (AAVSO), shown in Fig.~\ref{lightcurve}. We collected or inferred as many observational constraints as possible in order to limit viable parameter range for the models. 

\subsection{Observational constraints}
Several physical parameters are notoriously hard to derive for Mira stars. Spectroscopy has been hindered by the presence of many overlapping metallic and molecular lines and a lack of appropriate line lists to fit them, surface inhomogeneities, circumstellar envelopes, and extended, cool atmospheres. Moreover, the measurements are affected by the pulsation phase of the star. There have been successful attempts to derive abundances and [Fe/H] indices for Mira stars from infrared spectra, but no such observations exist for T~UMi \citep{uttenthaler-2015,dorazi-2018}.

Instead, we can infer some of the physical parameters with the help of Miras that are members of clusters. For those, the cluster ages and metallicities (or their observational proxies, the [Fe/H] indices) can be used as approximate values. The pulsation period of a Mira is a strong constraint on the average density and thus the evolution of the star, therefore a strong relationship between the periods and cluster ages is expected. It has also been shown that the kinematics of Miras correlate with their periods, indicating that a relation with the underlying metallicity or initial mass, or both, may exist \citep{feast-feh,kharcenko-2002}. This relation can be recovered from average cluster [Fe/H] indices and Mira member periods, and these can then be applied to other Mira stars.

\subsubsection{Age and\/ $[\mathrm{Fe}/\mathrm{H}]$}
An age relation was recently determined by \citet{mira-ages}, who collected Miras that belong to known star clusters both in the Milky Way and in the LMC. Using the original period, 313 d, and this relation, we estimate an age range between 0.5--5 Gyr. While this range is very broad, the lower limit rules out stars above 3~$\mathrm{M}_{\odot}$. The upper limit corresponds to masses around 1.2--1.3~$\mathrm{M}_\odot$, meaning that the 1.2~$\mathrm{M}_\odot$ model preferred by \citet{fadeyev2018}  is already bordering this constraint.

%Concerning the [Fe/H] index, 
\citet{feast-feh} collected average [Fe/H] indices of globular clusters with Miras and semiregular variables in them. Based on their relation, the 313 d period suggests a near-solar value of $[\mathrm{Fe}/\mathrm{H}]=-0.07$. However, given the large scatter of points and relative weakness of this constraint, we elected to use the solar value, corresponding to a metal abundance of Z = 0.014, following \citet{Asplund09}, for our detailed analysis, though we include a few test cases at slightly higher and lower metallicities.

\subsubsection{Chemical signatures, evolutionary state}
\label{spectroscopy}
High-resolution, optical-band spectra of the star were collected by \citet{tumi-uttenthaler2011} in 2009, with the HERMES spectrograph on the 1.2\,m Mercator telescope at La Palma. The observations suggest that the star is an oxygen-rich AGB star, and the C/O ratio is clearly below 1.0. The authors looked specifically for the signatures of only two elements, Tc and Li. The radioactive element Tc is produced in the \textit{s}-process, and it is brought to the surface via 3DU. Repeated dredge-ups also bring C to the surface, turning the star into a C-rich Mira, but the presence of Tc would signal the 3DU well before the C/O ratio exceeds 1.0. No signs of Tc were detected, indicating that the star has not yet undergone 3DU events, in accordance with being an O-rich Mira. 

The lack of detectable Tc in the spectrum could be indicative of substantial mass loss from T~UMi. A dichotomy exists among Miras, with Tc-poor stars showing higher levels of dust production than the Tc-rich ones. However, T~UMi was among the least reddened Tc-poor Miras, with a $K-[22] = 1.224$ infrared color index. This suggests that it is not experiencing strong dust mass loss \citep{uttenthaler-massloss,uttenthaler-massloss-2}.

Li is usually quickly destroyed in stars, but it can also be produced in AGB stars via the Cameron--Fowler mechanism \citep{Cameron-Fowler}. Li production can either be sustained in high-mass stars (over 4 $\mathrm{M}_\odot$) via hot-bottom burning and subsequently mixed into the cooler layers and eventually to the surface, or it can occur episodically during thermal pulses in lower-mass stars and then dredged to the surface \citep{Karakas-2010}.  
No Li was observed in T~UMi, with an upper limit of $\log\epsilon \lesssim 0.0$, i.e., $\log(N_\mathrm{Li}/N_\mathrm{H}) \lesssim -12$. The lack of any Li signature again suggests that T~UMi is a low-mass star (from this observation: $< 4 \mathrm{M}_\odot$), and that it has not yet undergone a 3DU event. 

\subsubsection{Gaia DR2 data and parallax estimates}
The star is present in the Gaia DR2 catalog, but was not identified as a variable, and the reported $G = 12.95$ mag brightness appears too low \citep{gaiadr2}. 
Closer inspection of the Gaia data revealed that the epoch photometry was affected by a faint outlier point, known to happen to a small number of other Gaia targets with large intrinsic variability \citep{arenou2018}.
Since this brightness value is subsequently used in the calibration of the astrometric data, the associated the parallax of T~UMi (\texttt{dr2.parallax} = $0.29 \pm 0.09$ mas) is also very likely to be inaccurate in Gaia DR2. 

However, we can give an estimate for the parallax of T~UMi using the PL relation. Some time-resolved $K$--band photometry of the star exists in the Caltech Two-Micron Sky Survey (2MSS) \citep{tmss}. Folding the data with the Mira period, the dereddened, pulsation-averaged brightness of the star is $K = 2.6 \pm 0.3$~mag (using an $A^c = 0.3$~mag correction, based on \citet{s-f-2011}). The PL-relation of \citet{whitelock2008} suggests an $M_K$ absolute brightness between $-7.2$ and $-8.2$ mag. We compute a parallax of $\pi_{PL} = 0.9 \pm 0.3$~mas based on this brightness information. 

This parallax estimate is at least a factor of two larger than the Gaia DR2 value, indicating that the geometric parallax is inaccurate. However, as this value is based on the PL-relation, it cannot be used to verify independently the absolute brightness of T~UMi---we must wait until the next Gaia data release to deduce this.

\subsection{Light variations}
 In this paper we focus on the visual observations. Although individual estimates are less accurate than the $V$ photometric data, they provide better temporal coverage. The data are numerous enough that we can safely discard the upper-limit values and only keep the positive observations from the AAVSO visual estimates. In total, we removed 892 upper-limit values and retained 14030 data points, spanning from JD2416780.6 (27/10/1904) to JD2458453.2361 (30/10/2018), as shown in Fig.~\ref{lightcurve}. 

As with the period decrease, the pulsation amplitude decreases considerably. During the long-period phase, the peak-to-peak amplitude hovers between 4.5--6.0 mag, but it has since decreased to 1.5--2.0 mag. Since both the AAVSO and General Catalogue of Variable Stars \citep[GCVS,][]{gcvs} classification schemes set a lower amplitude limit of 2.5 mag for Miras, T~UMi has been technically classifiable as a semiregular variable for more than a decade.

\subsubsection{Cycle lengths}
\begin{figure}
	\includegraphics[width=\columnwidth]{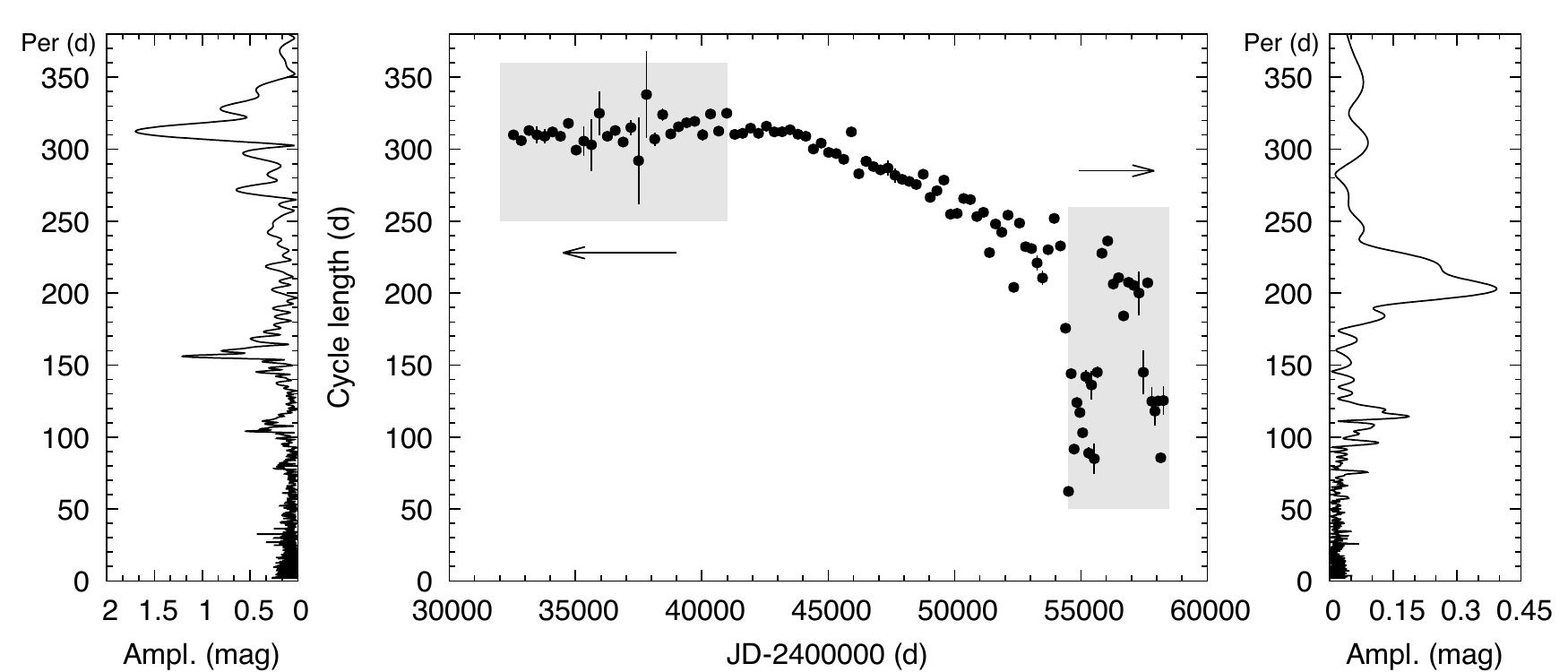}
    \caption{Middle: distances of successive (local) light maxima. Left and right: the corresponding periodograms from the early (grey-highlighted region in the upper left of middle panel) and late (grey-highlighted region in the lower right) sections of the light curve. The ``early'' region covers truncated JD 32000 to 41000, and the ``late'' region, 54500 to 58500.} 
    \label{cyclelength}
\end{figure}

In order to determine the current pulsation period of T~UMi, we first computed the individual cycle lengths, extending the work of \citet{tumi-szatmary2003}. We determined the times of maximum light for each pulsation cycle by fitting quadratic polynomials locally, and in cases where the data were too sparse, by visual estimates.
We expected either that the decreasing trend presented by \citet{tumi-szatmary2003} would continue, or that it would eventually stabilize at a new value. However, the recent observations delivered a more complex light curve than before, more reminiscent of beat, or double-mode, Cepheids than Miras. Our attempts to derive the cycle lengths resulted in a bifurcation of points that grouped around two different values. A comparison with the periodograms of light curve sections just before the decrease and afterwards, during the beating pattern (Fig.~\ref{cyclelength}), also indicates that the star is no longer pulsating in a single mode. This conclusion is in agreement with the suggestion made by \citet{foster2010}.

\begin{figure}
	\includegraphics[width=\columnwidth]{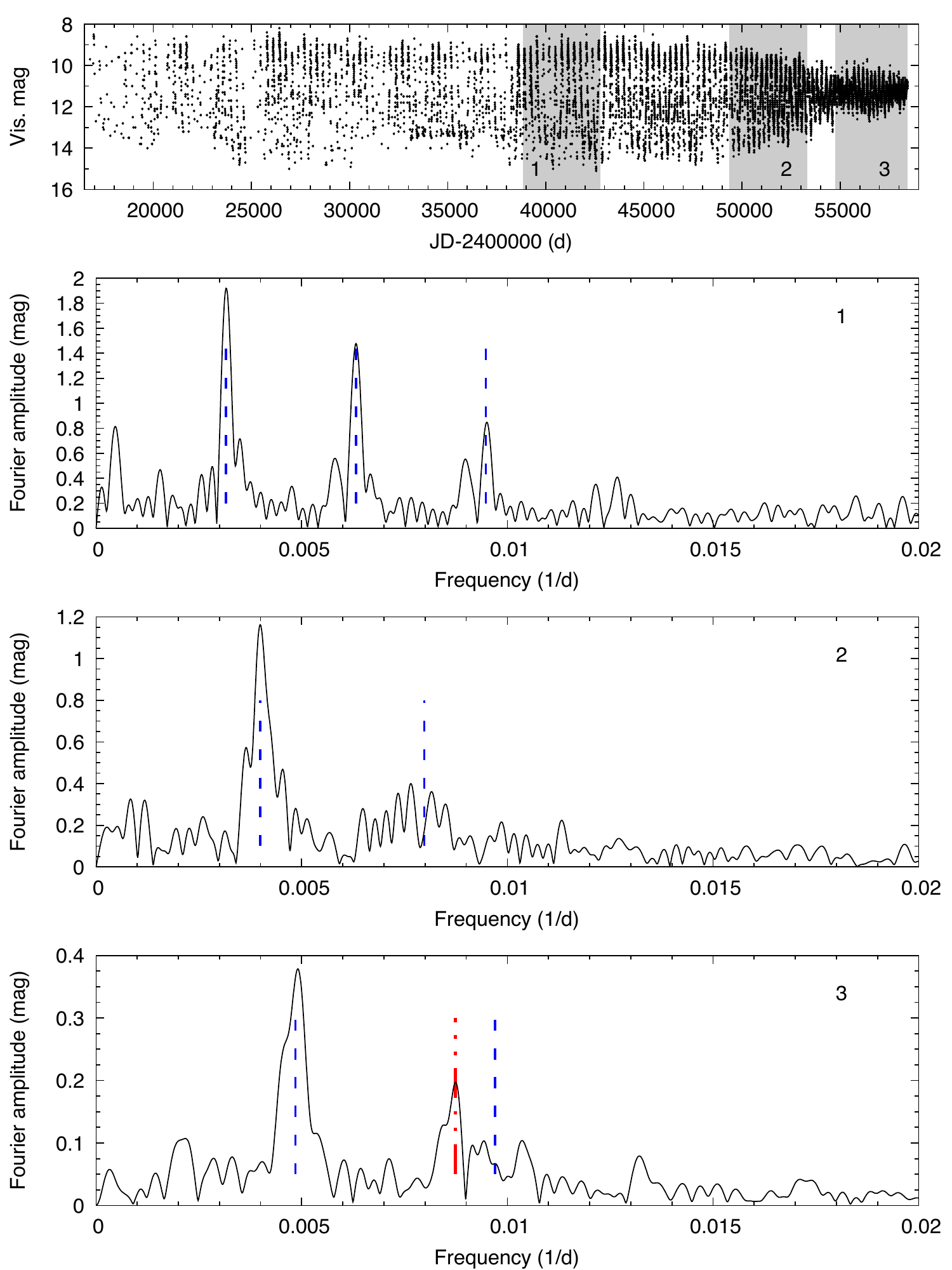}
    \caption{Top panel: the AAVSO visual data of T~UMi, with three different segments selected. Below: the Fourier spectra of each segment. Blue lines mark the dominant frequency and the positions of their harmonics at each panel. The red dot-dashed line at the bottom panel marks the new, independent frequency component.}
    \label{freqspectra}
\end{figure}

\subsubsection{Frequency spectra and time-frequency distributions}
The Fourier-spectrum of the complete data set shows a strong frequency peak at 0.0031966(3) d$^{-1}$ (or at a period of 312.8~d), bordered by a  broad forest of peaks towards higher frequencies and three additional harmonic peaks. This alone is indicative of the period change in the star. We found a mean pulsation frequency of 0.0031947(3) d$^{-1}$ ($P= 313.019$~d) before the inflection point.

We then split the light curve into smaller segments and calculated the spectra of each with Period04 \citep{period04}. The spectra clearly show the main period shortening towards the end of the data set. We show examples of the different segments in Fig.~\ref{freqspectra}. The three panels represent the pulsation of the star right before the period decrease started, during the decrease, and towards the end when the second mode emerged, with signal-to-noise ratio exceeding 5 over the residual spectrum.

A better method to follow the temporal evolution of the pulsation is to construct time-resolved maps of its frequency/period and amplitude contents. This was also done by \citet{tumi-szatmary2003}, whof showed that as the main pulsation frequency increased, the amplitude of its harmonics started to decrease, e.g., the cycles became more sinusoidal. 

We computed similar maps over the data using a sliding Gaussian window. We use 100 d timesteps and explore various Gaussians, settling eventually on 
a value of $\sigma = 1200$~d for the width of the window. We calculated the Lomb--Scargle periodograms for each timestep, and created a 2D map of the periodograms using the astropy package. 

\begin{figure}
	\includegraphics[width=\columnwidth]{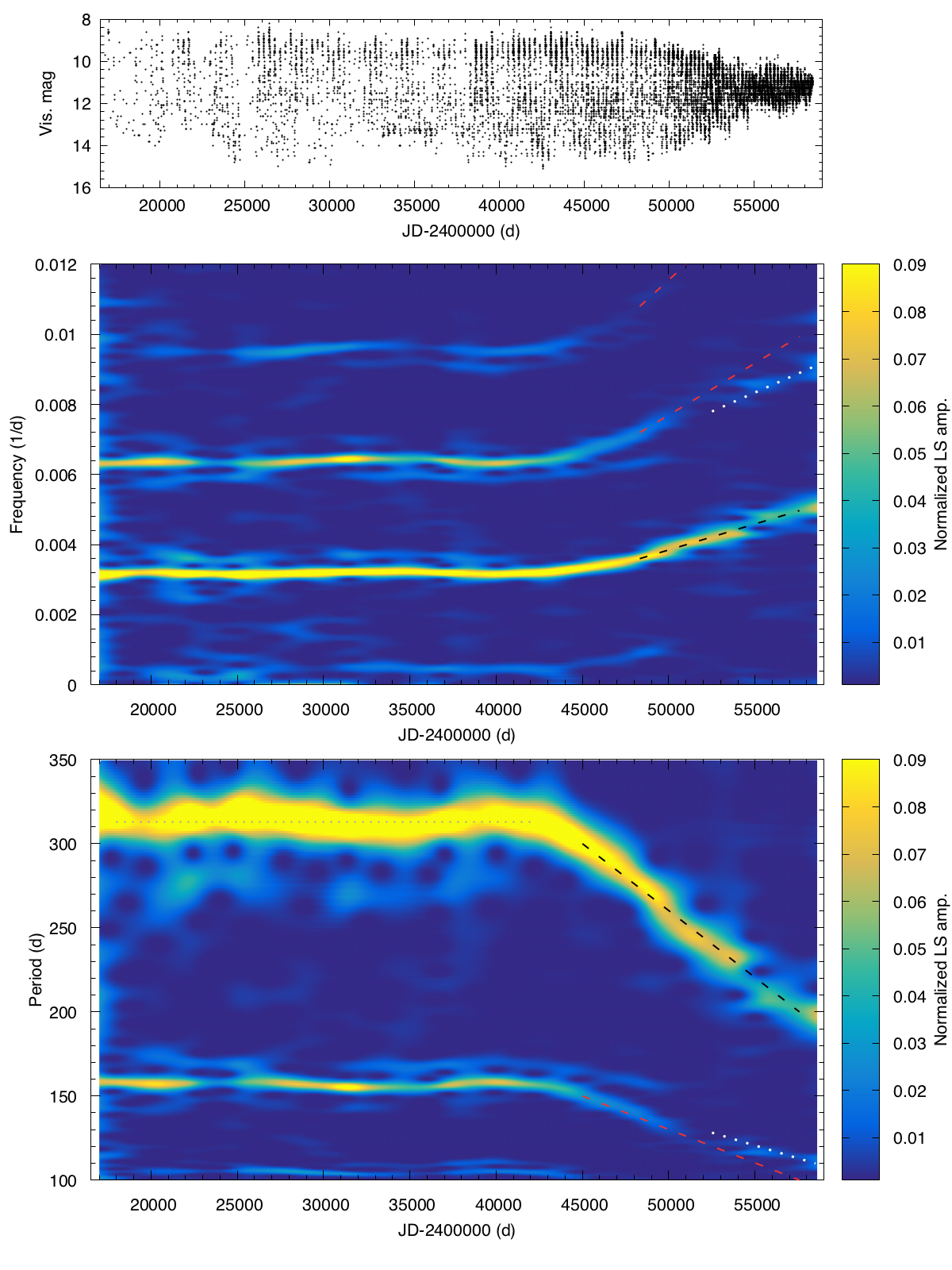}
    \caption{Top panel: the visual data of T~UMi. Middle panel: time-frequency distribution, bottom panel: time-period distribution. The main period is the strongest signal, the average, 313 d pulsation period is marked with a thin grey dotted line. The black and red dashed lines show the (actual or expected) position of the fundamental period and its harmonics. The white dotted line is the position of the new mode.}
    \label{tfd}
\end{figure}

Figure \ref{tfd} displays both the period-amplitude and frequency-amplitude distributions. Small changes are visible in the pulsation period, between 310--320 d, before the break---this is the meandering effect observed in many other Miras \citep{templeton2005}. This is then followed by a rapid decrease of nearly constant rate in period. The harmonic components of the pulsation period disappear, in agreement with earlier findings; this is apparent in the frequency plot (middle panel). Meanwhile, a new component is clearly visible at and after JD 2455000 in both the frequency and period visualizations, with an initial period of about 118 d that drops to 110 d by thxfe end of the data. The fundamental period of the star reaches 200 d by the end: with a period ratio of $P_1/P_0 \approx 0.55$ at that point, this second signal is clearly not the harmonic, but an independent pulsation mode. 

We compared the two pulsation periods towards the end of the observations to that of the multiperiodic semiregular variables \citep{kiss1999}. T~UMi fits perfectly to the most populated sequence of stars with period ratios near 0.55. This sequence corresponds to stars pulsating in the fundamental (FM) and first overtone (O1) radial modes, indicating that T~UMi has turned into an FM-O1 double-mode star.

\subsubsection{Pulsation-averaged brightness} 
The luminosity of a star entering a TP initially decreases as the H-burning shell is extinguished. This would be easily observable in the pulsation-averaged brightness of the star in infrared; however, no such observations have been taken since the period started to decrease. The visual light curve of Miras is harder to decipher.

The very large amplitudes (compared to those in the infrared) observed in the visual band are caused by metallic molecules, especially TiO, forming well above the photosphere when the star expands and cools in each pulsation cycle. These metallic oxides act as a ``sunscreen", blocking the outgoing visual photons coming from the stellar surface and re-emitting them at the top of the much cooler extended atmosphere, in the infrared. Therefore, one has to be cautious when interpreting changes in the average brightness of the star from the Mira to the semiregular phase, as changes to the pulsation amplitude could potentially affect the production of metallic oxides in the atmosphere.

\begin{figure}
	\includegraphics[width=\columnwidth]{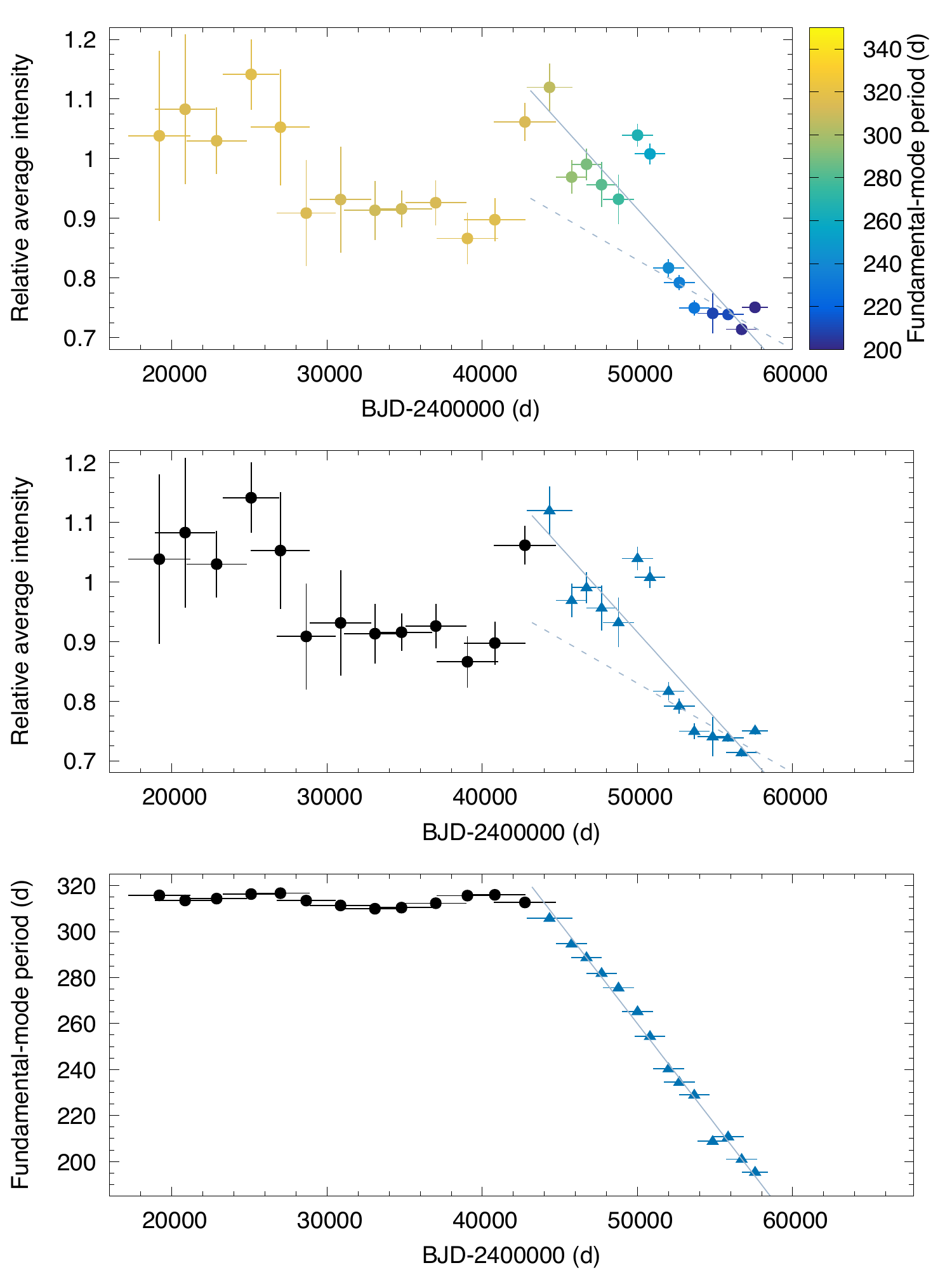}
    \caption{Top panel: change of pulsation-averaged relative intensity in T~UMi. Color indicates the dominant period of each segment. Middle and bottom panels: the intensity and period values separately. Black points refer to the time span before the rapid period change, blue triangles refer to the time span of rapid period change. The grey solid and dashed lines in the upper and middle panel show the rates of $-0.005$ and $-0.01$ yr$^{-1}$ in intensity, and the grey line in the lower panel is a $\dot P = -3.2$~d\,yr$^{-1}$ change period. }
    \label{intensity}
\end{figure}

A cursory investigation of the light curve suggests that the mean brightness has actually increased after the pulsation amplitude declined. 
However, the magnitude scale applies a non-linear transformation to the stellar flux. As \citet{tumi-szatmary2003} already noted, for such extreme variations, the average brightness in magnitude and the average flux converted to magnitudes are not only different, but also depend on the amplitude---this is due to the logarithmic scaling of the data. Therefore, converting the data back to physical units is more informative.

\citet{tumi-szatmary2003} also converted the light curve to intensity units, and they observed a small decrease in the cycle-averaged brightness of the star. Because the star transitioned to double-mode pulsation, we cannot simply extend the data forward using their method. Instead, we cut the intensity data into segments and fitted them individually in frequency space. Segment lengths are between 1000--4000 d, with longer bins where fewer data were recorded. This way, we were able to determine simultaneously the offset and the dominant frequency of each segment. We identified the meandering effect in the Mira phase observed by \citet{tumi-szatmary2003}, which was then followed by a clear downward trend, as seen in Fig.~\ref{intensity}.

Based on the visual data, the luminosity of the star decreased by an approximate rate of 1\% per year. However, the meandering nature of the light curve before that makes this estimate uncertain, and the rate could be as low as 0.5\%. Contemporaneously, the pulsation period was decreasing by $3.20\pm 0.15$~d/yr. 

Another important behavioral feature demonstrated by Fig.~\ref{intensity} is that while the pulsation-averaged brightness of the star changes considerably over the span of the data, the period only changes drastically in the second half. Some undulation is present in the period in the first half of the data as well, and it was noted by \citet{tumi-szatmary2003} that this correlates with the changes in average brightness. The amplitude, however, is much smaller. Whatever the cause behind the meandering in brightness and period, it has a much stronger effect on the luminosity of the star than its pulsation period, whereas the opposite is true for the onset of the thermal pulse. This suggests that the meandering must be connected to physical processes that only alter the outer layers of the star (and hence the brightness of the star), but which have lesser impact on the star's internal structure.

\section{Calculations}
\label{sec:calc}
We use MESA (Modules for Experiments in Stellar Astrophysics, \citealt{MESAIV})  version 10398 to generate evolutionary and structural models of T~UMi with varying initial mass estimates. We use GYRE version 5.1 \citep{GYRE} to construct oscillation frequency spectra from high-resolution structural profiles output at critical evolutionary timestamps, selected with high precision via a convergence algorithm of the authors' own construction.

\subsection{Model grid design}
The parameter space for our models was selected largely based on a soft interpretation of the classical observational limits. 
Our mass grid was chosen to encompass comfortably the best-fitting mass estimate for T~UMi from \citet{fadeyev2018} and to extend into a regime likely to be ruled out by evolutionary constraints (i.e., nuclear production signatures of higher-mass stars). This resulted in our consideration of models with initial masses between $1.0\text{--}3.0\, \mathrm{M}_\odot$, though we also investigated a model with an initial mass of $4.0\, \mathrm{M}_\odot$ to confirm that it produced evolutionary features inconsistent with T~UMi. 
We chose to adopt the solar metallicity of \citet{Asplund09} (corresponding to $Z_{\text{in}}=0.014$) for all formal analysis due to the weakness of the spectroscopic constraints and to provide easier comparison with similar calculations in the literature. However, we also computed coarse grids adopting each of $Z \in \{0.011,0.017,0.02 \}$ over the same mass range.
While the exact values of the initial Helium abundance ($Y$) and convective mixing length ($\alpha_{\text{MLT}}$) most appropriate for this star are also unlikely to be identical to the solar values \citep{Joyce2018a,Joyce2018b}, we are interested primarily in exploring the impact of initial mass. Hence $Y=0.27$ and $\alpha_{\text{MLT}}=2.0$ are global choices for our model grid as well.

We found that changes in metallicity primarily impacted the age at which the TP-AGB phase and 3DU (if present) activate, but did not impact the pre-3DU period ratios. It is likely that rerunning the complete analysis using a significantly higher or lower metallicity would change our global age estimate (with lower $Z$ resulting in older ages, and vice versa), however, the impact on the asteroseismic calculations is negligible for metallicity differences within the range of reasonable assumptions for T~UMi \citep{Joyce2018a}.
 
We adopt prescriptions for the boundaries of the Hydrogen- and Helium-burning regions fit to AGB stars in previous studies \citep{Tashibu}. 
These specifications were selected, in part, to increase the sensitivity of the models to the 3DU, because we want to ensure that we only consider pulses from models that do not exhibit the structural features of a Carbon star.
We use the ``cno\_extras\_o18\_to\_mg26.net'' nuclear reaction network as given in MESA version 10398, which includes all reactions required by the proton-proton chains, CNO cycle, Helium burning, and the additional isotopes which allow us to probe the production of Li$^6$, Li$^7$,  and Tc \citep{MESAIV}.

The Bl{\"o}cker cool wind AGB scheme is employed, with a scaling factor of $\eta_{\text{Bl\"ocker}}=0.1$ \citep{Blocker}. This value is chosen in accordance with a ``typical value'' estimated by \citet{Blocker}, and it is roughly the median among recent choices in the literature for similar models \citep{VenturaKarakas,Pignatari,Choi,Tashibu}. 
We performed a cursory exploration of the impact of an extreme change in the mass-loss rate by constructing coarse grids with $\eta_{\text{Bl\"ocker}}=0.5$ spanning four values of $Z$ and the standard mass grid. This change to $\eta_{\text{Bl\"ocker}}$ is found to lower the number of pulses considerably (e.g., from ${\sim}25$ to ${\sim}11$ for a model with initial mass $2.0\, \mathrm{M}_\odot$), but it does not appear to impact heavily the seismic features of structural models extracted during the relevant times in these sparser pulse spectra.

Checks against previous TP-AGB calculations in the literature were performed. These include comparisons between the number of thermal pulses produced for various masses and comparison of the conditions under which the 3DU occurred \citep{KarakasLat2007}. Our models were broadly consistent with the literature on both accounts, producing, e.g., roughly 20--25 thermal pulses before shell exhaustion for a star with initial mass $2.0$ $\mathrm{M}_{\odot}$ \citep{VenturaKarakas,Pignatari,Choi,Cristallo2015,Tashibu,KarakasLat2007,Blocker,colibri2019}.

\begin{figure}
	\includegraphics[width=\columnwidth]{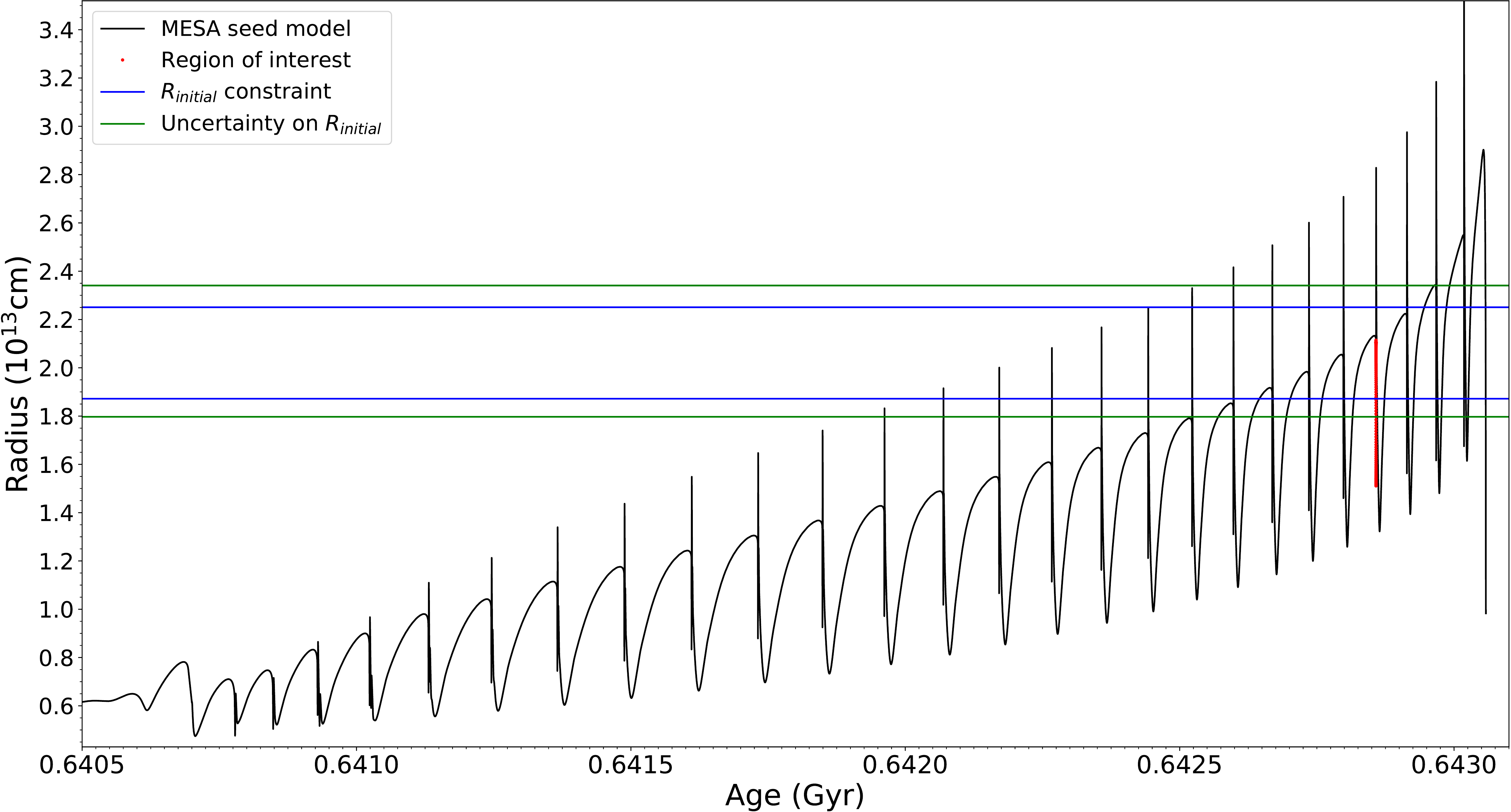}
	\caption{The full pulse spectrum for a model with initial mass 2.6 $\mathrm{M}_{\odot}$ is shown. Blue horizontal lines indicate the observational radial constraint appropriate for this model, and green horizontal lines indicate the observational uncertainty. The region highlighted in red shows one of several downward radial transition phases for which we must generate densely sampled grids of structural models in order to track the evolution of the star's frequency profile. }
    \label{region_of_interest}
\end{figure}

\begin{figure}
	\includegraphics[width=\columnwidth]{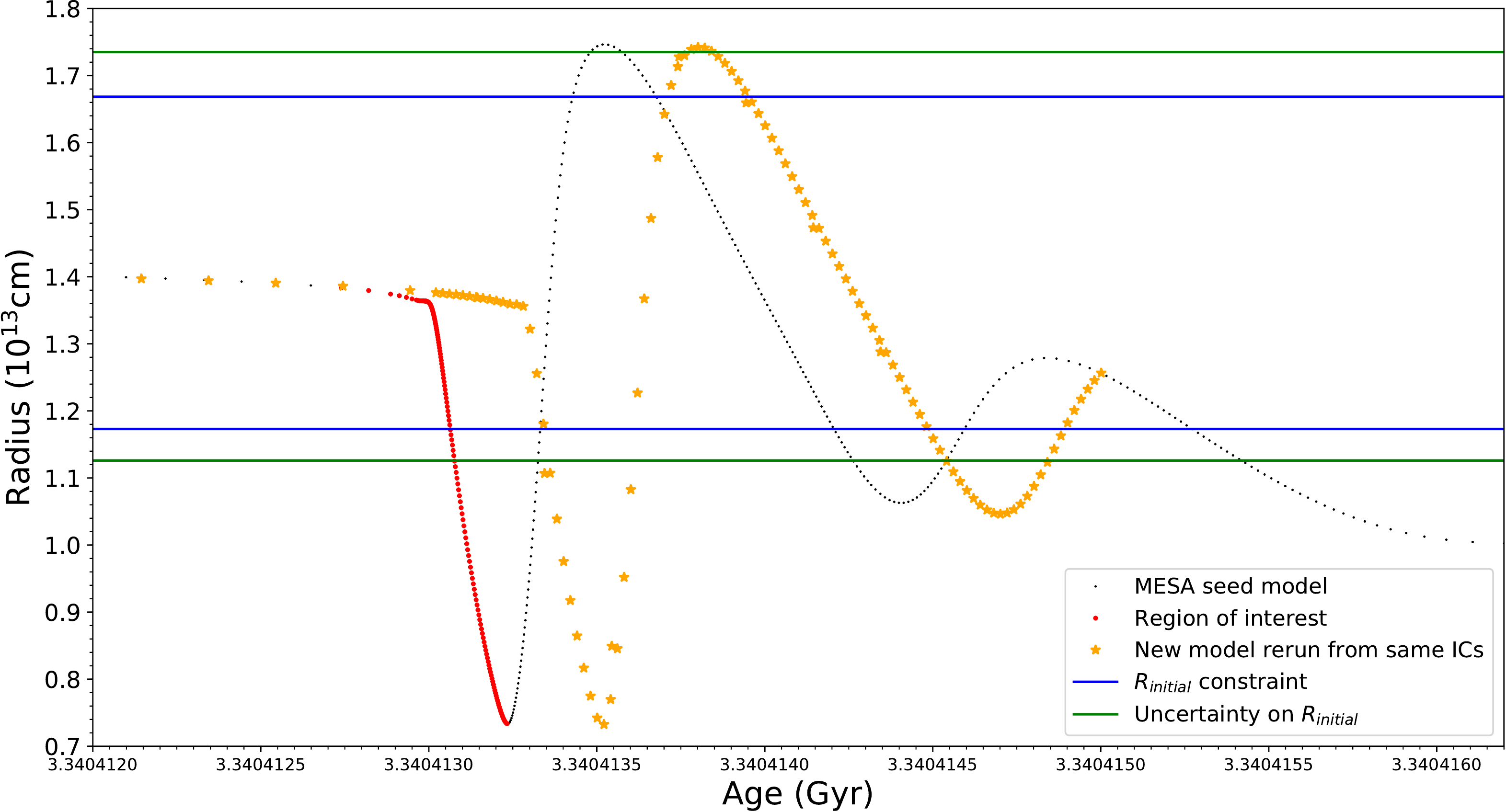}
	\caption{The impact of restart effects is demonstrated for a model with initial mass $1.4 \mathrm{M}_{\odot}$. The black curve line represents the seed profile from which timestamp estimates are drawn. The yellow curve represents a model run from the same initial conditions as the seed mode. Blue and green horizontal lines are as in Figure \ref{region_of_interest}. }
    \label{NLD}
\end{figure}

\subsection{Modeling procedure}
Due to the parameter sensitivity and extremely short evolutionary duration of the TP-AGB, standard grid modeling methods are poorly suited to this problem. 
The primary barrier is the difficulty of reproducing, exactly, a particular pulse profile despite identical initial conditions---a complication that arises due to the sensitivity of the evolution in this regime to the choice of timestep, which may deviate from previous timesteps in order to reproduce a model at the exact age needed. 
We refer to this issue subsequently as ``restart effects.''
Despite this obstacle, extreme precision is required in selecting the temporal regime associated with the appropriate radial transition. In some cases, it is necessary to sample regions of the pulse at a frequency of 5 yr, requiring our estimation of the timestamps to have a precision of $1$ in $10^9$.  
Offsets due to restart effects, as shown in Fig.~\ref{NLD}, can be three orders of magnitude larger than the temporal precision required. To overcome these difficulties, we instead implement a convergence scheme to obtain our seismic profiles. 

We begin by generating a set of lower (temporal) resolution evolutionary tracks ranging in initial mass (i.e., zero-age main sequence (ZAMS) mass) from 1.0 to 3.0 $\mathrm{M}_{\odot}$ in increments of 0.1 $\mathrm{M}_{\odot}$.
We first evolve each track to the onset of the TP phase (if it is activated), at which point we save both the structural model, call this \mod/, and the history of the secular evolution of the global state variables thus far. Using \mod/ as the initial conditions (ICs), we then evolve the model to the exhaustion of its envelope at the end of the TP phase. We concatenate the resulting evolutionary trajectory with the initial model and refer to this as a ``seed'' model. One such seed is built for each mass (and/or metallicity and $\eta_{\text{Bl\"ocker}}$ combination, where explored).

\begin{figure}
	\includegraphics[width=\columnwidth]{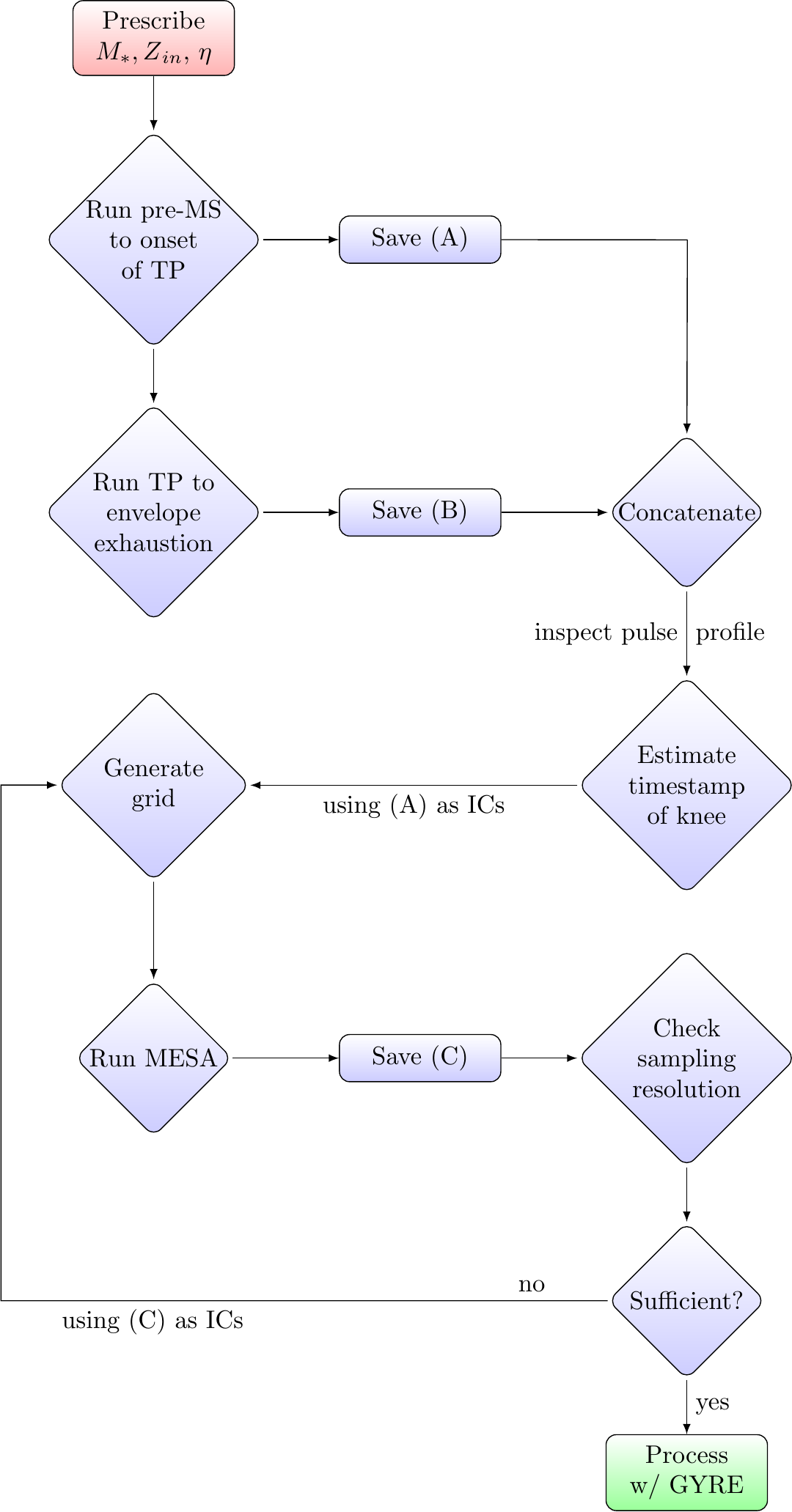}
        \caption{A flow of control diagram describing our modeling technique.} 
    \label{flow_of_control}
\end{figure}

The seed models are used to guide first estimates of the evolutionary timestamps at which the radial transition occurs. Using an interactive visualization tool developed by the authors, the timestamps of the points of interest---i.e., the ``knees'' of the thermal pulses, corresponding to the onset of downward inflection in period---are visually estimated. 
These timestamp estimates are then used as evolutionary stop conditions, where the first ICs are provided by \mod/ and subsequent ICs for a model $n$ are provided by a structural model saved at timestamp $n-1$. At each timestamp, we extract a high-resolution structural model from which the asteroseismic information can be computed with the GYRE \citep{GYRE}. A diagram describing this algorithm is provided in Figure~\ref{flow_of_control}.

This incremental approach mitigates the compounding of offsets caused by restart effects and preserves the structural continuity from one model to the next as best as possible. Additionally, the number of structural models from which to source ICs increases with number of trials, meaning convergence to the desired timestamp occurs more quickly with each iteration. Adopting an initial sampling rate of $dt=10,000$ years, sampling the desired radial transitions at a resolution of $5$ years typically takes 3 to 4 iterations. This scheme is several orders of magnitude faster and considerably more precise than launching grids of models with the same sampling resolution outright.

\begin{figure}
    \includegraphics[width=\columnwidth]{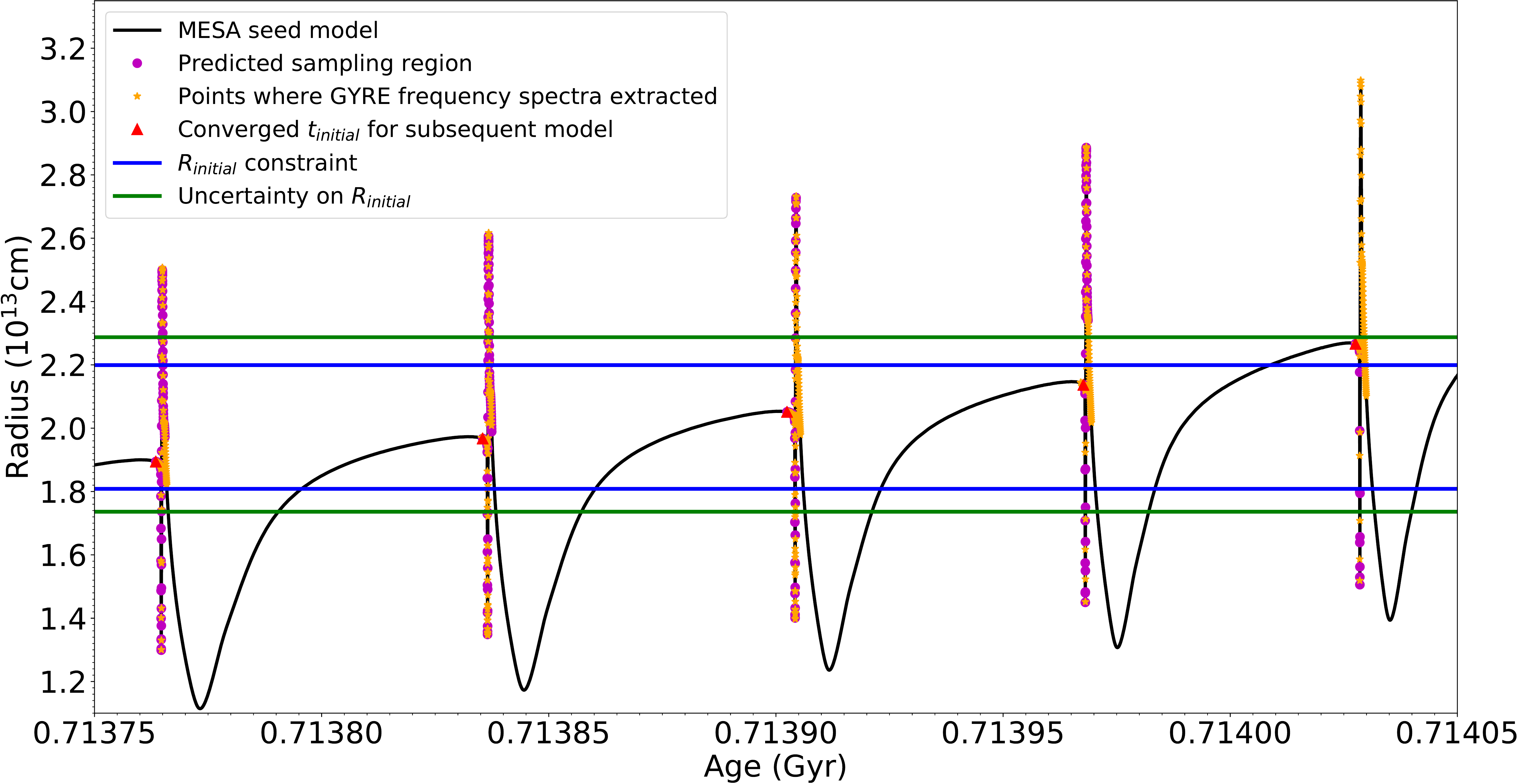}
	\caption{Predicted (purple) and actual (yellow) sampling locations for each viable pulsation are shown against the seed profile for a model with initial mass $2.5 \mathrm{M}_{\odot}$. Blue and green horizontal lines are radial constraints, as in previous figures.}
    \label{diagram}
\end{figure}

Our investigations are restricted to those pulses which satisfy observational constraints on the radius, as shown in Figure \ref{diagram}. The radial constraints are iteratively refined based on the radius--pulsation period relations determined for each mass. 
To ensure our results are robust across the changes in structure that occur with each pulse, we sample every pulse emerging in the permitted radial band with at least 20-year resolution.

For each mass, we recorded the pulse index of every pulse that occurred in the desired radial band. Across the grid, the models tended to reach the vicinity of the desired initial periods towards the end of the TP-AGB. For the lowest masses, this corresponded to roughly the $6^{\text{th}}\text{--}8^{\text{th}}$ pulses out of 10. At the high-mass end, the desired initial periods hit our radial limits around the $20^{\text{th}}$ to $24^{\text{th}}$ pulse out of $25\text{--}30$. These pulse numbers are in agreement with the findings of \citet{colibri2019}, who found that the concurrent excitation of the fundamental and first-overtone modes is most likely to happen in the latter-most TPs.

The main caveat to our technique is that no particular pulse necessarily reproduces the star perfectly, though modeling all pulses emerging within the permitted radial band mitigates this issue to some degree. We must also be clear that, though we find a general preference for higher pulse indices in our modeling, the evolutionary timestamp of the $n^{\text{th}}$ pulse and the total number of pulses in a pulse spectrum are extremely sensitive to the choices made for convective modeling parameters. None of these parameters (e.g. $\eta_{\text{Bl\"ocker}}, \alpha_{\text{MLT}}$, $\alpha_{\text{ovs}}$, various convective boundary sensitivities) is sufficiently well-calibrated to allow us to present, with confidence, a best-fitting pulse index. Rather, we can only ensure that the pulses we model do not correspond to profiles that contain post-3DU abundance signatures. As the convective modeling choices are largely ad-hoc, whereas the spectra are physical, we have made sure to prioritize consistency with our spectroscopic constraints.

GYRE can calculate both radial and non-radial modes, but we store only the adiabatic, linear periods of low-order radial modes, as we are interested in following the evolution of the fundamental mode and the first overtone specifically. We use GYRE version 5.1 to compute all of our final results, but we examined adiabatic frequency spectra generated with versions 4.2 and 5.2 as well. Results were consistent across these versions.

\subsection{Caveats to Modeling Technique}
During the course of this research, version 11532 of the MESA code was released. A major update was the integration of the Warsaw non-linear stellar pulsation code \citep{warsawcode-2008} into MESA as the new Radial Stellar Pulsations module \citep{mesa2019}.
We elected not to rerun our model grids using MESA version 11532 for two reasons. First, the Warsaw pulsation code was developed for Cepheid and RR Lyrae stars, not for pulsators with high $L/M$ ratios and/or very extended, cool outer envelopes, as it has been shown to encounter computational problems in the more luminous range of the type II Cepheid family \citep{smolec-2016}. Therefore, it is not well-suited for Mira and semiregular stars.

Moreover, multimode pulsations are both notoriously difficult and very time-consuming to model in the non-linear regime. Advances have been made for RR Lyrae and Cepheid stars with the Florida-Budapest pulsation code, which uses slightly different physical prescriptions than the Warsaw code \citep{kollath-2002,szabo-2004}. However, the accuracy of the physical processes in these 1D models, and by extension, their ability to precisely model double-mode stars, are still not settled conclusively \citep{smolec-moskalik-2008}. 

We thus do not attempt to model the pulsations of T~UMi in the non-linear regime. Instead, we focus on identifying the stellar parameter ranges and evolutionary stages where these pulsation periods are reflected in the linear regime. 

%%%%%%%%%%%%%%%%%%%%%%%%%%%%%%%%%%%%%%%%%%%%%%%%%%%%%%%%%%%%%%%%%%%%%%%%%%%%

A drawback of limiting our study to linear periods is that it lessens the accuracy with which we can identify certain fundamental parameters of the star.
While linear periods are a good approximation for small-amplitude variations, large-amplitude pulsations change the internal structure of the star and cause the nonlinear (e.g., observed) periods to shift. 
This was shown in a series of models by \citet{yaari1996}, who found that their nonlinear periods decreased by about 25\% compared to the linear periods at around 300 d. 
However, more recent studies have also shown that the degree of period shift depends on the physical approximations used in the models, especially the way convective energy transport is handled. For instance, models using MLT show larger shifts than those using time-dependent convection theory \citep{olivier-wood-2006}. Moreover, the same model produced a substantial period decrease for low-mass Mira models (0.6--0.9 $\mathrm{M}_\odot$, $-23$\%), but a clear period increase for higher-mass Mira models (1.32 and 1.43 $\mathrm{M}_\odot$, +8 and +13\%), suggesting that the envelope mass could be a deciding factor in the direction of the period shift \citep{lebzelter-wood-2005,wood-2007,kamath-2010,ireland-2011}. 

Based on these results, one can conclude that the periods are expected to increase by up to 15\% in the non-linear regime for stars above 1.1--1.3 $\mathrm{M}_\odot$---i.e., most of the synthetic stars we consider in this study.
We restrict our range of suitable initial linear, fundamental-mode period values to 270--310 d. Since the nonlinear period shift decreases as the pulsation amplitude declines, we defined a target range of $200\pm20$~d FM period and an O1/FM period ratio of $0.55\pm 0.02$ for the double-mode state.

\section{Results}
\label{sec:results}

The timescale resolution achieved in the calculations enables us, for the first time, to follow the temporal evolution of a star with decadal or better resolution as it enters a thermal pulse. We can hence provide analysis on how the pulsation periods change over time in unprecedented detail, and predict with fidelity what will happen to the star within a few decades.

\subsection{Confirmation of the ongoing TP}
The respective asteroseismic properties of the star and models are the main drivers for comparison.
In principle, the luminosity changes computed in the models can also be compared to the changes in average intensity observed in the star. We find general agreement between the rates in the models and in T~UMi, but the observations are too uncertain to be an especially rigid test of agreement, and the periods offer much stricter constraints. Moreover, the visual band only captures a fraction of the luminosity, so further follow-up in near-infrared bands would be required for a more instructive comparison. 

\begin{figure}
	\includegraphics[width=\columnwidth]{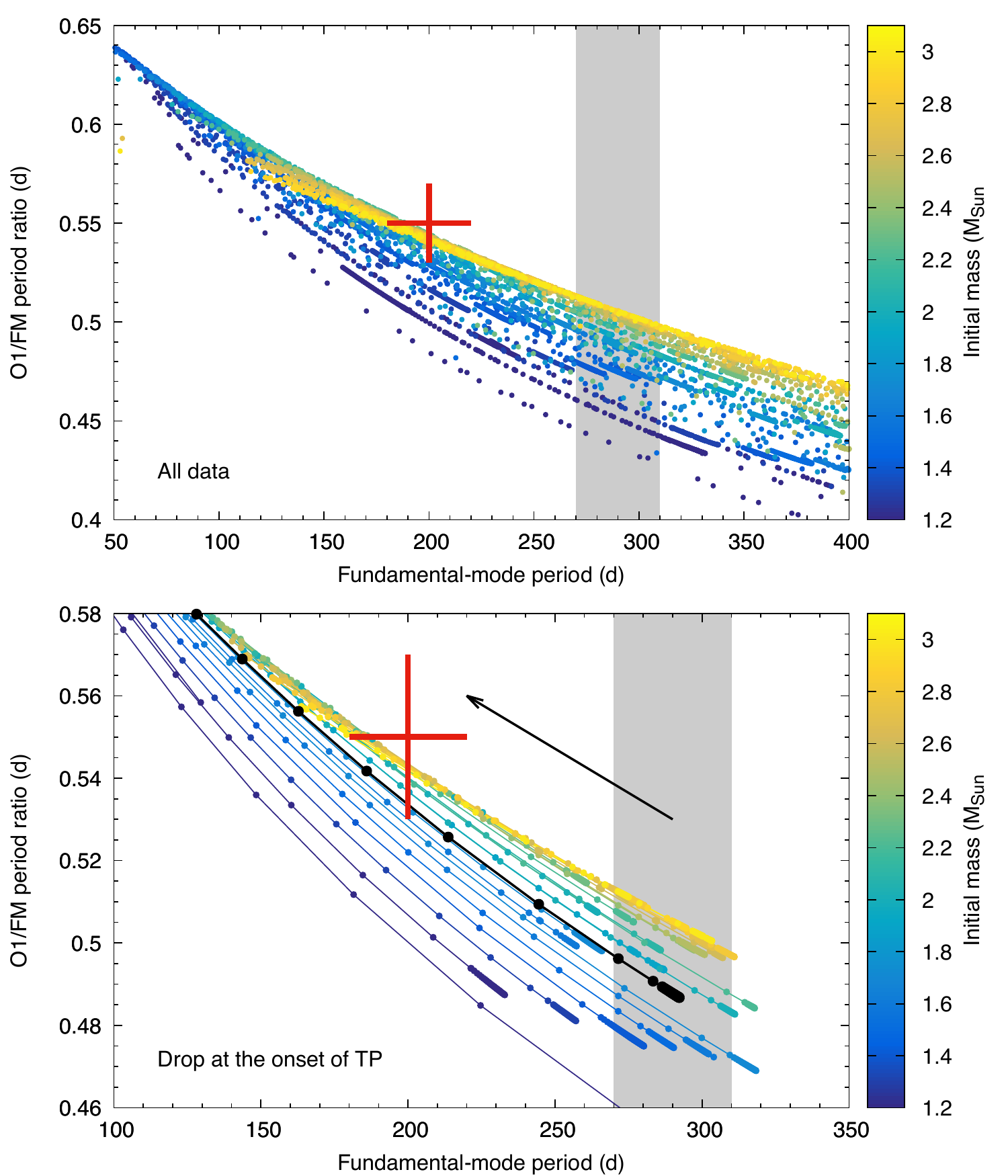}
	\caption{Upper panel: period vs period ratio values for all GYRE time steps. The grey band is the initial period range, the red cross marks the allowed FM period and period ratio  in the double-mode phase. Lower panel: same as above, but only showing the segments for the initial drop in period at the beginning of the TP, with the arrow indicating the direction of change. The black line highlights the 1.7~$\mathrm{M}_\odot$ model, the first that clearly traverses the cross.  }
    \label{palinkanet}
\end{figure}

We tracked all pulses that start from or near the $270-310$~d initial period range to see if they approach the correct parameter range at the double-mode state during the contraction phase. The latter criterion requires that the models
reach the expected 0.55 FM-O1 period ratio when the FM period drops to $\approx200$~d. 
Period ratios are plotted against the FM periods of the models in Fig.~\ref{palinkanet}, with the colors indicating the initial mass of the models, and the grey band and the red cross marking the Mira period range and the double-mode state, respectively. The upper panel shows all time steps for which frequency spectra were calculated with GYRE, the bottom panel only shows the decline sections at the start of the selected TPs. 

Figure \ref{palinkanet} very clearly demonstrates the advantage of observing two independent pulsation modes in the star. Earlier works assumed that the star was switching pulsation modes, but such a transition is possible without the onset of a TP. We show instead that the fundamental mode has been present in the star the whole time, but has recently experienced a rapid and massive reduction in period that can only be explained by the onset of a TP. The changes in physical parameters were sufficient to excite the first overtone as well, yielding two separate modes through which to trace the evolution of the star.

\begin{figure}
	\includegraphics[width=\columnwidth]{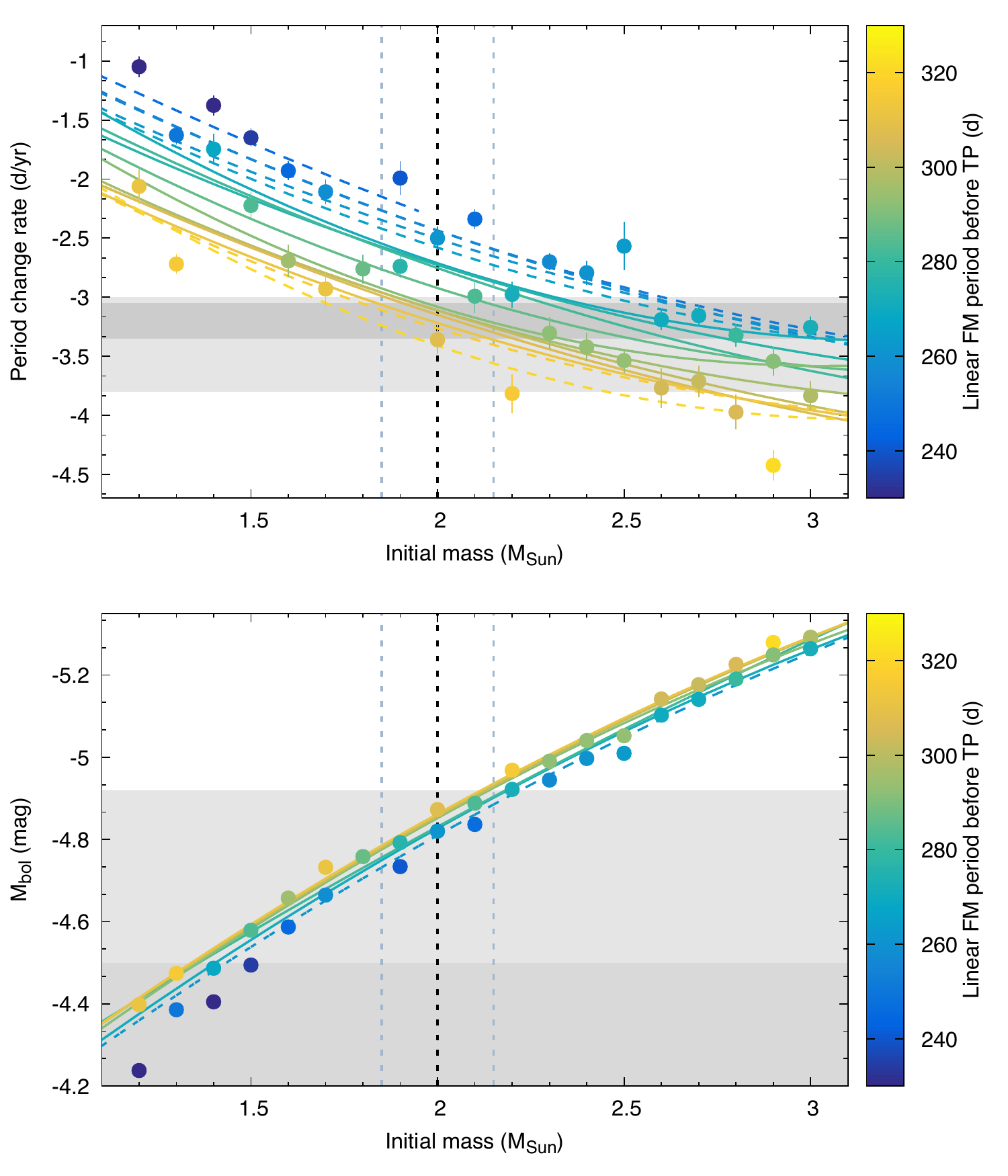}
    \caption{Period change rate and bolometric magnitude values from the various TP models. Solid lines are interpolated values for 10 d period increments. Grey shaded areas are observational constraints. Black and grey dashed lines mark the center and the edges of the best-fitting mass range.}
    \label{constraints}
\end{figure}

\subsection{Mass determination}
The two period values alone rule out the models in the low-mass regime, particularly the model with initial mass 1.2 $\mathrm{M}_\odot$ selected by \citet{fadeyev2018}. In the $1.2 \mathrm{M}_{\odot}$  case, when the first overtone reaches the 114~d value that \citet{fadeyev2018} used as constraint, the period of the fundamental mode is 234~d, i.e., considerably longer than the observed value. Only models with initial masses above 1.6 $\mathrm{M}_\odot$ start to approach the desired period ratio, and models above 2.0 $\mathrm{M}_\odot$ give the best fit to the observations. In the upper mass regime, the models are less sensitive to the period ratio at around the 200~d period, and the models start to overlap there in Fig.~\ref{palinkanet}. In this mass range, the slopes of the fundamental period curves begin to change instead, reaching lower period ratios as they shrink. This manifests as an inversion of the color coding in Fig.~\ref{palinkanet}, as the curves reorder in the 100--150 d period range. 

The age constraint based on the cluster Mira results rule out models that reach the TP-AGB before 0.5 Gyr or after 5 Gyr, meaning that the 3.0~$\mathrm{M}_\odot$ and 1.2~$\mathrm{M}_\odot$ models are disfavored by this metric alone. However, this constraint is weak relative to information derived from the pulsations: the initial period, the periods in the double-mode state, and the rate of period change of the fundamental mode. Hence, these edge cases are considered.

\subsubsection{Period change rate limits}
The models sample the drop in period after the knee well enough to determine $\dot P$ for the fundamental mode, and so synthetic $\dot P$ can be compared to the measurements of $\dot P$ in T~UMi. 
Two issues here are that (1) we cannot identify a specific pulse index for T~UMi, as this is highly contingent on modeling choices for convective parameters, and (2) that pulse calculations may align such that they border, but miss, the desired initial period range entirely. Therefore, we compute both $\dot P$ and the time it takes for the period to reach its minimum value for the two pulses nearest to the desired initial period range, and then use these data to interpolate relations at constant periods.

The upper panel of Fig.~\ref{constraints} shows the model-derived $\dot P$ values against the values determined by \citet{tumi-szatmary2003} (light grey) and this work (dark grey). This comparison places the lower limit again at 1.8--1.9 $\mathrm{M}_\odot$ for the initial mass of the star, but leaves the upper limit somewhat uncertain by this metric.

\begin{figure}
	\includegraphics[width=\columnwidth]{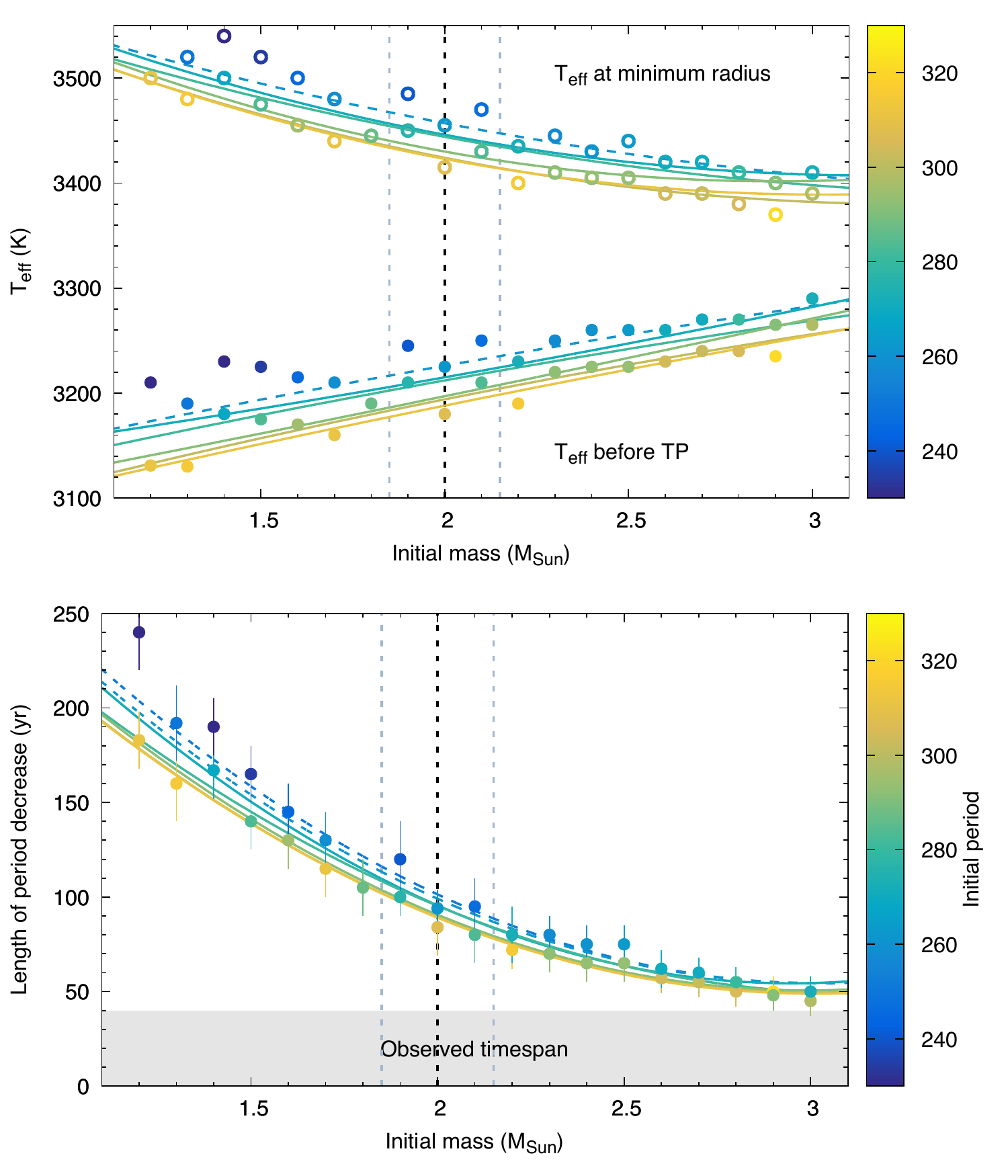}
    \caption{Effective temperature values based on the models, and length of the period decrease. Nomenclature is the same as in Fig.~\ref{constraints}. }
    \label{teff-bounce}
\end{figure}

\subsubsection{Luminosity limits}
The asteroseismic constraints do not provide a strong upper limit for the initial mass of T~UMi. We can, however, compare the bolometric luminosities of the models against bolometric absolute magnitudes derived for Mira variables. Calculating $\mathrm{M}_\mathrm{bol}$ is not a straightforward task for these stars, but \citet{Guandalini2008} and \citet{whitelock-2012} derived P-$\mathrm{M}_\mathrm{bol}$ relations based on various Mira and semiregular pulsators both in the Milky Way and other Local Group galaxies.  
These relations include various uncertainties, ranging from different stellar populations in each galaxy to the exact choice of bolometric corrections. 
The P-$\mathrm{M}_\mathrm{bol}$ relation of \citet{whitelock-2012} provides a faint estimate reaching $\mathrm{M}_\mathrm{bol} = -4.5$~mag that covers only the lowest masses, and it is mutually exclusive with the mass range allowed the by $\dot P$ constraint. Another relation derived by \citet{Guandalini2008} gives approximately $\mathrm{M}_\mathrm{bol} = -4.9$~mag; we use the latter as the upper limit for the bolometric brightness.
Incorporating these uncertainties, this comparison suggests that the initial mass of the star did not exceed  2.2~$\mathrm{M}_\odot$.

Synthesizing these constraints, we report an initial mass of 
$2.0\pm0.15$~$\mathrm{M}_\odot$
for T~UMi. This constitutes the first ever high-accuracy stellar mass determination based on an analysis of the properties of a thermal pulse. 
We also report a model-derived age of $1.17 \pm 0.21$ Gyr,
a value consistent with and considerably more precise than the age range deduced observationally. 

The individual pulses---within a given TP-AGB pulse spectrum---that fit best are those whose initial radii correspond to linear periods between 290 and 310~d; however, these are identified based on models that invoke a solar chemical composition ($Z=0.014$) and a specific prescription for the convective modelling parameters, all of which are invariant in this analysis. Hence, small changes in our mass estimate could occur if the true composition or interior physics of the star differ significantly from this.
Fundamental parameters derived in this work are indicated as such in Table \ref{fundpar}.

\subsection{Radius and effective temperature inferences}
With these mass limits at hand, we can examine other physical properties of the relevant models to provide ranges for additional fundamental stellar parameters: the radius and $T_{\rm eff}$. 
The pre-transition values of both parameters and the degree of change induced by the TP depend on the mass, where the lower-mass models shrink and heat up considerably more than those of higher mass. If we consider only the preferred mass range, we find that the radius of the star was $290\pm 15$~R$_\odot$ in the Mira phase, and the minimum value it should reach in the near future is $180\pm 15$~R$_\odot$---a reduction of about 40\%. 
 
The T$_\mathrm{eff}$ values at the beginning of the TP increase slightly with mass, but change by less than 150~K over the whole range. The models which best reproduce the observational constraints provide an effective temperature of T$_\mathrm{eff} = 3200 \pm 30$~K for the pre-TP Mira phase. The expected maximum is T$_\mathrm{eff}= 3440 \pm 30$~K. Estimates of effective temperature, however, are particularly sensitive to the modeling parameters (especially those that control aspects of heat transfer), and these uncertainties are only appropriate for the particular choices made in our grid.

\begin{figure}
	\includegraphics[width=\columnwidth]{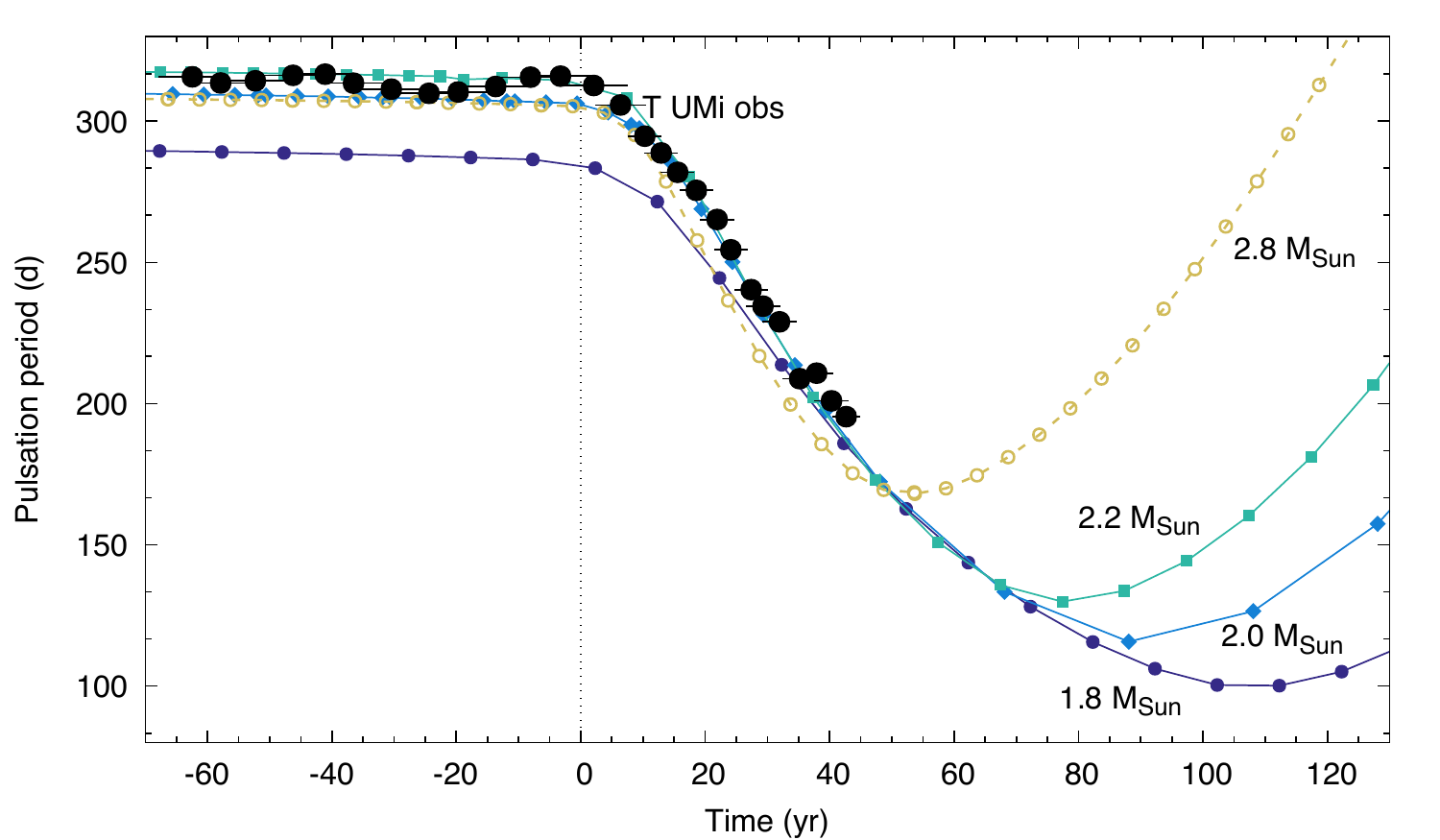}
    \caption{Comparison between the ongoing period decrease in T~UMi, and the same in three models with different initial masses. Depending on the mass, the pulsation period may be expected to increase again in a few decades.
    }
    \label{prediction}
\end{figure}

\subsection{Additional physical parameters and predictions}

The dimming and shrinking of the star should soon reverse. Exactly when, though, is a function of both stellar mass and the thermal pulse index: the higher the mass or the later the pulse, the faster the upturn in radius occurs. 
Given the lack of double-mode, non-linear Mira pulsation models, we cannot predict which pulsation state the star will occupy at that time. It could remain in the double-mode state, but it could also transition to pure, first-overtone pulsation before reaching the minimum radius.
Irrespective of which mode or modes are excited, the pulsation periods strongly depend on the radius, and future observations will provide the indications required to follow the evolutionary trajectory of T~UMi. In addition, timing the mode transitions will provide important insights into pulsation mode selection in AGB stars.

In the bottom panel of Fig.~\ref{teff-bounce}, we plot the time spans in the models between the knee and the time of shortest period, with the gray dashed line indicating the time that has already passed for T~UMi. If the initial mass of the star is small---around 1.5~$\mathrm{M}_\odot$---it could take another century to reach the minimum period. If the star is in the high-mass regime---above 2.5~$\mathrm{M}_\odot$---we will see the pulsation period level off and begin its ascent within as little as 10--30 years.

In the model-preferred mass range, the reversal of the period decrease happens in about $50\pm10$ years. Slowdown of the decrease could become noticeable about a decade before that. This means that we will be able to constrain further the mass of the star simply by continuing observation over the next few decades. As such, a verification or refutation of our predictions is accessible within our lifetime.

A few examples of the projected evolution of the FM period in the models, compared to that of T~UMi, are shown in Fig.~\ref{prediction}.

\begin{figure}
	\includegraphics[width=\columnwidth]{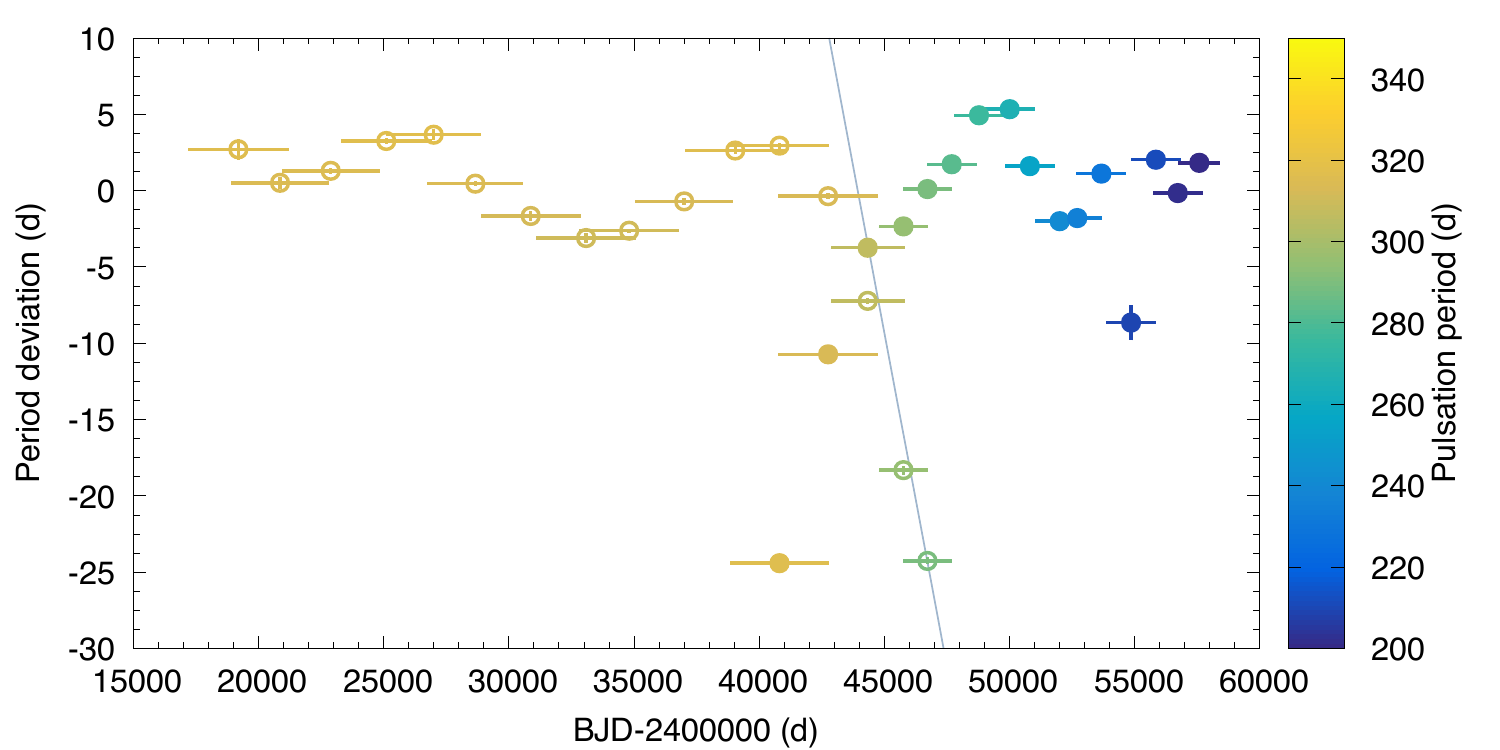}
    \caption{Meandering, or cyclic period undulations in the pulsation. Empty circles: the same as in Fig.~\ref{intensity} but with the 313~d average period subtracted. Filled circles: same, but with the linear period decrease term subtracted. Colors indicate the pulsation period. The small variations clearly continue in the second half of the data, but the cycle length is shortened. }
    \label{undulation}
\end{figure}

\subsection{Non-evolutionary effects}
The models predict a smooth and monotonic change in period (and surface luminosity) until the start of the~TP, followed by a much more rapid, but still monotonic, decrease after the knee. However, the models do not predict the meandering effect we see in Mira stars, suggesting that meandering is not a known evolutionary effect.
The average intensity data suggest that T~UMi is not fading at a constant rate, but the pulsation periods give a more conclusive indication: subtracting the average value, 313~d, from the first portion of the data, and a linear fit with a constant $\dot P$ from the latter portion (e.g., post-knee) results in Fig.~\ref{undulation}. 
Clearly, the meandering phenomenon persists into the TP phase of the star, but the cycle length seems to shorten. The origin of these slow variations is not well understood, but one possible explanation is thermal oscillations in the envelope \citep{templeton2005}. In that case, however, the cycle length would be defined by the Kelvin--Helmholtz timescale, which increases in duration as the luminosity and radius drops---the opposite of what we observe here.

%%%%%%%%%%%%%%%%%%%%% table %%%%%%%%%%%%%%%%%%%%%%%

\begin{deluxetable*}{l r l}
\tablecaption{Summary of Classical, Seismic, and Modelled Parameters of T~UMi. \label{fundpar}}
\tablecolumns{3}
%\tablewidth{0pt}
\tablehead{
\colhead{Parameter} &
\colhead{Value} &
\colhead{Inference Method}
}
\startdata
%%%%%%%% observationally inferred seismic parameters
Pulsation period before knee (d)     & 313.0 & light curve analysis, this work \\
Present-day fundamental period (d)     & $198.1\pm1.0$ & light curve analysis, this work  \\
Present-day overtone period (d)     & $111.0\pm0.5$ & light curve analysis, this work  \\
$\dot P$ (d/yr) 		& $-3.8 \pm 0.4$ & light curve analysis$^1$ \\
$\dot P$ (d/yr) & $-3.20 \pm 0.15$~ & light curve analysis, this work\\
	Period ratio $P_1/P_0$ &  ${\approx} 0.55$ & this work \\
%
%%%%%%%% observationally inferred classical parameters
Parallax (mas)                     & $0.9\pm0.3$ & period--luminosity relation \\ 
Approx.\ [Fe/H] (dex)                        & $-0.07$ & cluster period--[Fe/H] relation$^2$  \\
$K$ brightness before knee (mag)         & $2.6 \pm 0.3$ & 2MSS$^3$ (dereddened, $A^c = 0.3$ mag)$^4$ \\
%
%%%%%%%%% model-derived fundamental parameters
	ZAMS Mass ($\mathrm{M}_{\odot}$) 			& $2.0\pm0.15$   & model grid \\
	Present-day Mass ($\mathrm{M}_{\odot}$)      & $1.66 \pm 0.10$ & model grid \\
Radius ($R_{\odot}$)            	& $290\pm 15$   & model grid \\
T$_{\text{eff}}$ (K)  				& $3200 \pm 30$ &  model grid \\
Age (Gyr)							& $1.17 \pm 0.21$   & model grid \\
	Pulse index						    & $12\text{--}15$       & model grid \\
%
%%%%%%%%%%%% adopted parameters in model grid
$Z$  				                & $0.014\pm0.003$ &  estimate based on $Z_{\odot}$ and spectroscopy \\
$\alpha_{\text{MLT}}$ & 2.0 & adopted MESA solar--calibrated value \\
$\eta_{\text{Bl\"ocker}}$ & 0.1 &  literature estimates$^{+}$\\ 
\hline
\enddata
\tablecomments{In order: observationally inferred seismic parameters, observationally inferred classical parameters, model-derived fundamental parameters, and parameters adopted in the model grid are listed.
\\  Uncertainties quoted for the model-inferred properties do not include individual uncertainties in $Z$, $\alpha_{\text{MLT}}$, $\eta_{\text{Bl\"ocker}}$, or other prescribed parameters, but rather take into account extremal values of the 1$\sigma$ uncertainty on the best-fitting mass estimate. 
\\$^1${\citet{tumi-szatmary2003}}, $^2${\citet{feast-feh}},  $^3${\citet{tmss}}, $^4${\citet{s-f-2011}}
}
\tablenotetext{+}{Informed by literature values and optimized to achieve pulse numbers consistent with results of other modeling groups \citep{VenturaKarakas,Pignatari,Choi,Cristallo2015,Tashibu,KarakasLat2007,Blocker}.}
\end{deluxetable*}
%%%%%%%%%%%%%%%%%%%%% table %%%%%%%%%%%%%%%%%%%%%%%

\section{Conclusions and future work}
\label{sec:summary}
The thermally pulsing AGB phase affords a rare opportunity to observe an individual star's evolution in a human lifetime. In this study, we investigated the pulsations of a former single Mira star, T~UMi, that transitioned into a semiregular variable over the last few decades to test whether this transition is due to the recent onset of a thermal pulse.
We found that the period of the fundamental-mode radial pulsation has been shortening steadily in the last 40 years. The appearance of a second mode---the first overtone---and transition to double--mode pulsation rules out mode switching as a possible cause of the observed changes. 

Through sophisticated analysis and the synthesis of decades of observations, we have concluded that the dimming and shrinking of T~UMi should soon reverse. Exactly when, though, is a function of both stellar mass and the thermal pulse index: the higher the mass or the later the pulse, the faster the reversal of the current radial decline will occur. 

We calculated theoretical evolutionary models and stellar oscillation spectra with MESA and GYRE, respectively, to test whether the changes to the pulsation properties of T~UMi---and the initial reduction in stellar radius associated with it---are caused by the onset of a TP. We developed a sophisticated adaptive sampling scheme to select high-precision time stamps from an initial evolutionary model.

The isolation of the desired TPs was complicated by the emergence of restart effects that could shift the starting time of a TP by up to several hundred years when launching subsequent iterations of the models from the same initial timestamp. The convergence algorithm mitigated these effects, allowing us to compute spectra of linear pulsation periods over the TPs with a temporal resolution of 1 part in $10^9$. We used this to track changes in the pulsation periods occurring on the timescale of 5--20 y.

We have derived various constraints from the observations: the initial pulsation period in the Mira phase, the periods of the two modes during the semiregular phase, and the rate of period change of the fundamental mode. We compared these to the calculated TPs in the model grid to infer the fundamental properties of the star, particularly the initial mass value.
One drawback of our analysis, however, is the use of linear pulsation models to compute the synthetic asteroseismic parameters.

The asteroseismic parameters alone provide excellent constraints for the initial mass of T~UMi. We are able to rule out conclusively the earlier 1.2 $\mathrm{M}_\odot$ value reported by \citet{fadeyev2018}. We then compared the synthetic luminosities to the P--M$_\mathrm{bol}$ relation derived by \citet{whitelock-2012} and \citet{Guandalini2008} to derive an upper limit for the initial mass, although this constraint is comparatively softer.

By combining these constraints, we report an optimal initial mass of $\mathrm{M}_\mathrm{ZAMS} = 2.0\pm0.15\,\mathrm{M}_\odot$ for T~UMi. This is the first time that the mass of a single TP-AGB star has been derived in a comprehensive way, using considerably more detailed asteroseismic and classical constraints than any previous attempt.

From the model grid, we infer additional physical parameters characterizing T~UMi as it was before the onset of the pulse. These include an age of $1.17 \pm 0.21$ Gyr, an effective temperature of $T_\mathrm{eff} = 3200 \pm 30$, and a radius of $R = 290\pm 15$~R$_\odot$.
For the present-day (i.e., mid-TP) mass, we obtain $M = 1.66 \pm 0.10$~M$_\odot$, but this is heavily dependent on the choice of model parameters, especially on the mass-loss rate and convective mixing prescription (\citealt{Joyce2018b}). The derived and inferred fundamental parameters are summarized in Table \ref{fundpar}.

We predict that the decrease in T~UMi's period will continue for another five decades, with changes in $\dot P$ becoming noticeable several years to a decade prior to its transition. Tracking these changes will require sustained observations both from successive generations of amateur astronomers and small-aperture professional instruments scanning the sky repeatedly (e.g., the Evrysope, MASCARA, or Fly's Eye arrays,  \citet{evryscope-2015,mascara-2012,flyseye-2013}).

A TP-induced transition likewise offers the opportunity to study the pulsations of AGB stars directly. Timing the mode transitions can provide important insights into pulsation mode selection in AGB stars. Moreover, since the star is currently shrinking and heating up, but will eventually expand and cool down, the TP will provide directional information on the blue- and redward evolution \citep{szabo-2004}.

The confirmation of the TP-induced luminosity decrease requires long-term pulsation-averaged brightness measurements for the star. Near-infrared photometry will be crucial for following the evolution of the star through the TP, since energy output of an AGB star peaks in those bands.
Moreover, updated geometric parallaxes in future Gaia releases will allow us to improve our absolute brightness estimate of T~UMi. More accurate knowledge of the period change rates and the bolometric magnitude of the star will then allow us to pinpoint the physical parameters even more precisely.

Finally, the combination of slow science, the long--term and very productive pro--am collaboration that has been achieved by the observer network of the AAVSO, and state-of-the-art numerical stellar modeling can be applied to other stars as well. Multiple other pulsating variable stars are suspected to be in some sub-phase of a TP, or transitioning from O-rich to C-rich stars. Such candidates are excellent contenders for similar studies, as we anticipate the science to follow from T~UMi's continuation along the thermal pulse we are currently witnessing.

\vspace{5mm}

\facilities{AAVSO}

\software{MESA \citep{MESAIV}, GYRE \citep{GYRE}, astropy \citep{astropy:2013,astropy:2018}, matplotlib \citep{matplotlib}, numpy \citep{numpy}, scipy \citep{scipy}, Period04 \citep{period04}, gnuplot}

\acknowledgements
\section*{Acknowledgements}
Authors Moln\'ar and Joyce contributed equally to this manuscript. Fruitful discussions with Maria Lugaro, Carolyn Doherty, and Emese Plachy are gratefully acknowledged. L. M. was supported by the Premium Postdoctoral Research Program of the Hungarian Academy of Sciences. The research leading to these results has received funding from the LP2014-17 and LP2018-7 Lend\"ulet grants of the Hungarian Academy of Sciences and from the NKFIH K-115709 grant of the National Research, Development and Innovation Office of Hungary. We acknowledge with thanks the variable star observations from the AAVSO International Database contributed by observers worldwide and used in this research. This research has made use of the SIMBAD database, operated at CDS, Strasbourg, France, and NASA's Astrophysics Data System Bibliographic Services.

M. J. was supported by Martin Asplund, the Research School of Astronomy and Astrophysics at Australian National University, and funding from Australian Research Council grant number DP150100250. M. J. wishes to acknowledge John Bourke for his assistance in revising this manuscript and Maria Lugaro for financial support of her scientific residences at Konkoly Observatory. 
% AGB stars are cool, coffee chats are acknowledged.

%%%%%%%%%%%%%%%%%%%% REFERENCES %%%%%%%%%%%%%%%%%%

\bibliography{all-11-20-18.bib} 

\begin{thebibliography}{}
\expandafter\ifx\csname natexlab\endcsname\relax\def\natexlab#1{#1}\fi
\providecommand{\url}[1]{\href{#1}{#1}}
\providecommand{\dodoi}[1]{doi:~\href{http://doi.org/#1}{\nolinkurl{#1}}}
\providecommand{\doeprint}[1]{\href{http://ascl.net/#1}{\nolinkurl{http://ascl.net/#1}}}
\providecommand{\doarXiv}[1]{\href{https://arxiv.org/abs/#1}{\nolinkurl{https://arxiv.org/abs/#1}}}

\bibitem[{{Arenou} {et~al.}(2018){Arenou}, {Luri}, {Babusiaux}, {Fabricius},
  {Helmi}, {Muraveva}, {Robin}, {Spoto}, {Vallenari}, {Antoja},
  {Cantat-Gaudin}, {Jordi}, {Leclerc}, {Reyl{\'e}}, {Romero-G{\'o}mez}, {Shih},
  {Soria}, {Barache}, {Bossini}, {Bragaglia}, {Breddels}, {Fabrizio},
  {Lambert}, {Marrese}, {Massari}, {Moitinho}, {Robichon}, {Ruiz-Dern},
  {Sordo}, {Veljanoski}, {Eyer}, {Jasniewicz}, {Pancino}, {Soubiran}, {Spagna},
  {Tanga}, {Turon}, \& {Zurbach}}]{arenou2018}
{Arenou}, F., {Luri}, X., {Babusiaux}, C., {et~al.} 2018, \aap, 616, A17,
  \dodoi{10.1051/0004-6361/201833234}

\bibitem[{{Asplund} {et~al.}(2009){Asplund}, {Grevesse}, {Sauval}, \&
  {Scott}}]{Asplund09}
{Asplund}, M., {Grevesse}, N., {Sauval}, A.~J., \& {Scott}, P. 2009, Ann. Rev.
  Astr. and Astroph., 47, 481, \dodoi{10.1146/annurev.astro.46.060407.145222}

\bibitem[{{Astropy Collaboration} {et~al.}(2013){Astropy Collaboration},
  {Robitaille}, {Tollerud}, {Greenfield}, {Droettboom}, {Bray}, {Aldcroft},
  {Davis}, {Ginsburg}, {Price-Whelan}, {Kerzendorf}, {Conley}, {Crighton},
  {Barbary}, {Muna}, {Ferguson}, {Grollier}, {Parikh}, {Nair}, {Unther},
  {Deil}, {Woillez}, {Conseil}, {Kramer}, {Turner}, {Singer}, {Fox}, {Weaver},
  {Zabalza}, {Edwards}, {Azalee Bostroem}, {Burke}, {Casey}, {Crawford},
  {Dencheva}, {Ely}, {Jenness}, {Labrie}, {Lim}, {Pierfederici}, {Pontzen},
  {Ptak}, {Refsdal}, {Servillat}, \& {Streicher}}]{astropy:2013}
{Astropy Collaboration}, {Robitaille}, T.~P., {Tollerud}, E.~J., {et~al.} 2013,
  \aap, 558, A33, \dodoi{10.1051/0004-6361/201322068}

\bibitem[{{Bedding} {et~al.}(1998){Bedding}, {Zijlstra}, {Jones}, \&
  {Foster}}]{rdor-1998}
{Bedding}, T.~R., {Zijlstra}, A.~A., {Jones}, A., \& {Foster}, G. 1998, MNRAS,
  301, 1073, \dodoi{10.1046/j.1365-8711.1998.02069.x}

\bibitem[{{Bl{\"o}cker}(1995)}]{Blocker}
{Bl{\"o}cker}, T. 1995, \aap, 297, 727

\bibitem[{{Buchler} {et~al.}(2004){Buchler}, {Koll{\'a}th}, \&
  {Cadmus}}]{buchler-2004}
{Buchler}, J.~R., {Koll{\'a}th}, Z., \& {Cadmus}, Robert~R., J. 2004, ApJ, 613,
  532, \dodoi{10.1086/422903}

\bibitem[{{Cameron} \& {Fowler}(1971)}]{Cameron-Fowler}
{Cameron}, A.~G.~W., \& {Fowler}, W.~A. 1971, ApJ, 164, 111,
  \dodoi{10.1086/150821}

\bibitem[{{Campbell} \& {Pickering}(1912)}]{Campbell-1912}
{Campbell}, L., \& {Pickering}, E.~C. 1912, Annals of Harvard College
  Observatory, 63

\bibitem[{{Choi} {et~al.}(2016){Choi}, {Dotter}, {Conroy}, {Cantiello},
  {Paxton}, \& {Johnson}}]{Choi}
{Choi}, J., {Dotter}, A., {Conroy}, C., {et~al.} 2016, ApJ, 823, 102,
  \dodoi{10.3847/0004-637X/823/2/102}

\bibitem[{{Cristallo} {et~al.}(2015){Cristallo}, {Straniero}, {Piersanti}, \&
  {Gobrecht}}]{Cristallo2015}
{Cristallo}, S., {Straniero}, O., {Piersanti}, L., \& {Gobrecht}, D. 2015,
  ApJS, 219, 40, \dodoi{10.1088/0067-0049/219/2/40}

\bibitem[{{Doherty} {et~al.}(2017){Doherty}, {Gil-Pons}, {Siess}, \&
  {Lattanzio}}]{doherty2017}
{Doherty}, C.~L., {Gil-Pons}, P., {Siess}, L., \& {Lattanzio}, J.~C. 2017,
  PASA, 34, e056, \dodoi{10.1017/pasa.2017.52}

\bibitem[{{D'Orazi} {et~al.}(2018){D'Orazi}, {Magurno}, {Bono}, {Matsunaga},
  {Braga}, {Elgueta}, {Fukue}, {Hamano}, {Inno}, {Kobayashi}, {Kondo},
  {Monelli}, {Nonino}, {Przybilla}, {Sameshima}, {Saviane}, {Taniguchi},
  {Thevenin}, {Urbaneja-Perez}, {Watase}, {Arai}, {Bergemann}, {Buonanno},
  {Dall'Ora}, {Da Silva}, {Fabrizio}, {Ferraro}, {Fiorentino}, {Francois},
  {Gilmozzi}, {Iannicola}, {Ikeda}, {Jian}, {Kawakita}, {Kudritzki}, {Lemasle},
  {Marengo}, {Marinoni}, {Mart{\'\i}nez-V{\'a}zquez}, {Minniti}, {Neeley},
  {Otsubo}, {Prieto}, {Proxauf}, {Romaniello}, {Sanna}, {Sneden}, {Takenaka},
  {Tsujimoto}, {Valenti}, {Yasui}, {Yoshikawa}, \& {Zoccali}}]{dorazi-2018}
{D'Orazi}, V., {Magurno}, D., {Bono}, G., {et~al.} 2018, ApJ, 855, L9,
  \dodoi{10.3847/2041-8213/aab100}

\bibitem[{{Fadeyev}(2018)}]{fadeyev2018}
{Fadeyev}, Y.~A. 2018, Astronomy Letters, 44, 546,
  \dodoi{10.1134/S1063773718070010}

\bibitem[{{Feast} \& {Whitelock}(2000)}]{feast-feh}
{Feast}, M., \& {Whitelock}, P. 2000, in Astrophysics and Space Science
  Library, ed. F.~{Matteucci} \& F.~{Giovannelli}, Vol. 255, 229

\bibitem[{{Foster}(2010)}]{foster2010}
{Foster}, G. 2010, Journal of the American Association of Variable Star
  Observers (JAAVSO), 38, 140

\bibitem[{{Gaia Collaboration} {et~al.}(2018){Gaia Collaboration}, {Brown},
  {Vallenari}, {Prusti}, {de Bruijne}, {Babusiaux}, {Bailer-Jones}, {Biermann},
  {Evans}, {Eyer}, {Jansen}, {Jordi}, {Klioner}, {Lammers}, {Lindegren},
  {Luri}, {Mignard}, {Panem}, {Pourbaix}, {Randich}, {Sartoretti}, {Siddiqui},
  {Soubiran}, {van Leeuwen}, {Walton}, {Arenou}, {Bastian}, {Cropper},
  {Drimmel}, {Katz}, {Lattanzi}, {Bakker}, {Cacciari}, {Casta{\~n}eda},
  {Chaoul}, {Cheek}, {De Angeli}, {Fabricius}, {Guerra}, {Holl}, {Masana},
  {Messineo}, {Mowlavi}, {Nienartowicz}, {Panuzzo}, {Portell}, {Riello},
  {Seabroke}, {Tanga}, {Th{\'e}venin}, {Gracia-Abril}, {Comoretto},
  {Garcia-Reinaldos}, \& {Teyssier}}]{gaiadr2}
{Gaia Collaboration}, {Brown}, A.~G.~A., {Vallenari}, A., {et~al.} 2018, \aap,
  616, A1, \dodoi{10.1051/0004-6361/201833051}

\bibitem[{{G\'al} \& {Szatm\'ary}(1995)}]{tumi-gal1995}
{G\'al}, J., \& {Szatm\'ary}, K. 1995, \aap, 297, 461

\bibitem[{{Grady} {et~al.}(2019){Grady}, {Belokurov}, \& {Evans}}]{mira-ages}
{Grady}, J., {Belokurov}, V., \& {Evans}, N.~W. 2019, MNRAS, 483, 3022,
  \dodoi{10.1093/mnras/sty3284}

\bibitem[{{Guandalini} \& {Busso}(2008)}]{Guandalini2008}
{Guandalini}, R., \& {Busso}, M. 2008, \aap, 488, 675,
  \dodoi{10.1051/0004-6361:200809932}

\bibitem[{Hunter(2007)}]{matplotlib}
Hunter, J.~D. 2007, Computing In Science \& Engineering, 9, 90,
  \dodoi{10.1109/MCSE.2007.55}

\bibitem[{{Ireland} {et~al.}(2011){Ireland}, {Scholz}, \&
  {Wood}}]{ireland-2011}
{Ireland}, M.~J., {Scholz}, M., \& {Wood}, P.~R. 2011, \mnras, 418, 114,
  \dodoi{10.1111/j.1365-2966.2011.19469.x}

\bibitem[{Jones {et~al.}(2001--)Jones, Oliphant, Peterson, {et~al.}}]{scipy}
Jones, E., Oliphant, T., Peterson, P., {et~al.} 2001--, {SciPy}: Open source
  scientific tools for {Python}.
\newblock \url{http://www.scipy.org/}

\bibitem[{{Joyce} \& {Chaboyer}(2018{\natexlab{a}})}]{Joyce2018a}
{Joyce}, M., \& {Chaboyer}, B. 2018{\natexlab{a}}, ApJ, 856, 10,
  \dodoi{10.3847/1538-4357/aab200}

\bibitem[{{Joyce} \& {Chaboyer}(2018{\natexlab{b}})}]{Joyce2018b}
---. 2018{\natexlab{b}}, ApJ, 864, 99, \dodoi{10.3847/1538-4357/aad464}

\bibitem[{{Kamath} {et~al.}(2010){Kamath}, {Wood}, {Soszy{\'n}ski}, \&
  {Lebzelter}}]{kamath-2010}
{Kamath}, D., {Wood}, P.~R., {Soszy{\'n}ski}, I., \& {Lebzelter}, T. 2010,
  MNRAS, 408, 522, \dodoi{10.1111/j.1365-2966.2010.17137.x}

\bibitem[{{Karakas} \& {Lattanzio}(2007)}]{KarakasLat2007}
{Karakas}, A., \& {Lattanzio}, J.~C. 2007, Publ. Astron. Soc. Aust., 24, 103,
  \dodoi{10.1071/AS07021}

\bibitem[{Karakas(2017)}]{Karakas2017}
Karakas, A.~I. 2017, "Low- and Intermediate-Mass Stars" (Cham: Springer
  International Publishing), 1--21.
\newblock \url{https://doi.org/10.1007/978-3-319-20794-0_117-1}

\bibitem[{{Karakas} {et~al.}(2010){Karakas}, {Campbell}, \&
  {Stancliffe}}]{Karakas-2010}
{Karakas}, A.~I., {Campbell}, S.~W., \& {Stancliffe}, R.~J. 2010, ApJ, 713,
  374, \dodoi{10.1088/0004-637X/713/1/374}

\bibitem[{{Kerschbaum} \& {Hron}(1992)}]{kerschbaum-1992}
{Kerschbaum}, F., \& {Hron}, J. 1992, \aap, 263, 97

\bibitem[{{Kharchenko} {et~al.}(2002){Kharchenko}, {Kilpio}, {Malkov}, \&
  {Schilbach}}]{kharcenko-2002}
{Kharchenko}, N., {Kilpio}, E., {Malkov}, O., \& {Schilbach}, E. 2002, \aap,
  384, 925, \dodoi{10.1051/0004-6361:20020084}

\bibitem[{{Kiss} {et~al.}(1999){Kiss}, {Szatm{\'a}ry}, {Cadmus}, \&
  {Mattei}}]{kiss1999}
{Kiss}, L.~L., {Szatm{\'a}ry}, K., {Cadmus}, R.~R., J., \& {Mattei}, J.~A.
  1999, \aap, 346, 542.
\newblock \doarXiv{astro-ph/9904128}

\bibitem[{{Kiss} {et~al.}(2000){Kiss}, {Szatm{\'a}ry}, {Szab{\'o}}, \&
  {Mattei}}]{kiss-2000}
{Kiss}, L.~L., {Szatm{\'a}ry}, K., {Szab{\'o}}, G., \& {Mattei}, J.~A. 2000,
  Astronomy and Astrophysics Supplement Series, 145, 283,
  \dodoi{10.1051/aas:2000353}

\bibitem[{{Koll{\'a}th} {et~al.}(2002){Koll{\'a}th}, {Buchler}, {Szab{\'o}}, \&
  {Csubry}}]{kollath-2002}
{Koll{\'a}th}, Z., {Buchler}, J.~R., {Szab{\'o}}, R., \& {Csubry}, Z. 2002,
  \aap, 385, 932, \dodoi{10.1051/0004-6361:20020182}

\bibitem[{{Lacour} {et~al.}(2009){Lacour}, {Thi{\'e}baut}, {Perrin}, {Meimon},
  {Haubois}, {Pedretti}, {Ridgway}, {Monnier}, {Berger}, {Schuller},
  {Woodruff}, {Poncelet}, {Le Coroller}, {Millan-Gabet}, {Lacasse}, \&
  {Traub}}]{Lacour2009}
{Lacour}, S., {Thi{\'e}baut}, E., {Perrin}, G., {et~al.} 2009, ApJ, 707, 632,
  \dodoi{10.1088/0004-637X/707/1/632}

\bibitem[{{Law} {et~al.}(2015){Law}, {Fors}, {Ratzloff}, {Wulfken},
  {Kavanaugh}, {Sitar}, {Pruett}, {Birchard}, {Barlow}, {Cannon}, {Cenko},
  {Dunlap}, {Kraus}, \& {Maccarone}}]{evryscope-2015}
{Law}, N.~M., {Fors}, O., {Ratzloff}, J., {et~al.} 2015, PASP, 127, 234,
  \dodoi{10.1086/680521}

\bibitem[{{Lebzelter} \& {Wood}(2005)}]{lebzelter-wood-2005}
{Lebzelter}, T., \& {Wood}, P.~R. 2005, \aap, 441, 1117,
  \dodoi{10.1051/0004-6361:20053464}

\bibitem[{{Lenz} \& {Breger}(2005)}]{period04}
{Lenz}, P., \& {Breger}, M. 2005, Communications in Asteroseismology, 146, 53,
  \dodoi{10.1553/cia146s53}

\bibitem[{{Mattei} \& {Foster}(1995)}]{tumi-mattei1995}
{Mattei}, J.~A., \& {Foster}, G. 1995, Journal of the American Association of
  Variable Star Observers (JAAVSO), 23, 106

\bibitem[{{Neugebauer} \& {Leighton}(1969)}]{tmss}
{Neugebauer}, G., \& {Leighton}, R.~B. 1969, {NASA SP}, Vol. 3047, {Two-micron
  sky survey. A preliminary catalogue} ({Washington}: NASA)

\bibitem[{{Olivier} \& {Wood}(2005)}]{olivier-wood-2006}
{Olivier}, E.~A., \& {Wood}, P.~R. 2005, MNRAS, 362, 1396,
  \dodoi{10.1111/j.1365-2966.2005.09414.x}

\bibitem[{{P{\'a}l} {et~al.}(2013){P{\'a}l}, {M{\'e}sz{\'a}ros},
  {Cs{\'e}p{\'a}ny}, {Jask{\'o}}, {Schlaffer}, {Vida}, {Mez{\H{o}}},
  {D{\"o}brentei}, {Farkas}, {Kiss}, {Ol{\'a}h}, \&
  {Reg{\'a}ly}}]{flyseye-2013}
{P{\'a}l}, A., {M{\'e}sz{\'a}ros}, L., {Cs{\'e}p{\'a}ny}, G., {et~al.} 2013,
  Astronomische Nachrichten, 334, 932, \dodoi{10.1002/asna.201211962}

\bibitem[{{Paxton} {et~al.}(2018){Paxton}, {Schwab}, {Bauer}, {Bildsten},
  {Blinnikov}, {Duffell}, {Farmer}, {Goldberg}, {Marchant}, {Sorokina},
  {Thoul}, {Townsend}, \& {Timmes}}]{MESAIV}
{Paxton}, B., {Schwab}, J., {Bauer}, E.~B., {et~al.} 2018, ApJS, 234, 34,
  \dodoi{10.3847/1538-4365/aaa5a8}

\bibitem[{{Paxton} {et~al.}(2019){Paxton}, {Smolec}, {Gautschy}, {Bildsten},
  {Cantiello}, {Dotter}, {Farmer}, {Goldberg}, {Jermyn}, {Kanbur}, {Marchant},
  {Schwab}, {Thoul}, {Townsend}, {Wolf}, {Zhang}, \& {Timmes}}]{mesa2019}
{Paxton}, B., {Smolec}, R., {Gautschy}, A., {et~al.} 2019, arXiv e-prints,
  arXiv:1903.01426.
\newblock \doarXiv{1903.01426}

\bibitem[{{Pignatari} {et~al.}(2016){Pignatari}, {Herwig}, {Hirschi},
  {Bennett}, {Rockefeller}, {Fryer}, {Timmes}, {Ritter}, {Heger}, {Jones},
  {Battino}, {Dotter}, {Trappitsch}, {Diehl}, {Frischknecht}, {Hungerford},
  {Magkotsios}, {Travaglio}, \& {Young}}]{Pignatari}
{Pignatari}, M., {Herwig}, F., {Hirschi}, R., {et~al.} 2016, ApJS, 225, 24,
  \dodoi{10.3847/0067-0049/225/2/24}

\bibitem[{{Price-Whelan} {et~al.}(2018){Price-Whelan}, {Sip{\H{o}}cz},
  {G{\"u}nther}, {Lim}, {Crawford}, {Conseil}, {Shupe}, {Craig}, {Dencheva},
  {Ginsburg}, {VanderPlas}, {Bradley}, {P{\'e}rez-Su{\'a}rez}, {de Val-Borro},
  {Paper Contributors}, {Aldcroft}, {Cruz}, {Robitaille}, {Tollerud},
  {Coordination Committee}, {Ardelean}, {Babej}, {Bach}, {Bachetti}, {Bakanov},
  {Bamford}, {Barentsen}, {Barmby}, {Baumbach}, {Berry}, {Biscani}, {Boquien},
  {Bostroem}, {Bouma}, {Brammer}, {Bray}, {Breytenbach}, {Buddelmeijer},
  {Burke}, {Calderone}, {Cano Rodr{\'\i}guez}, {Cara}, {Cardoso}, {Cheedella},
  {Copin}, {Corrales}, {Crichton}, {D{\textquoteright}Avella}, {Deil},
  {Depagne}, {Dietrich}, {Donath}, {Droettboom}, {Earl}, {Erben}, {Fabbro},
  {Ferreira}, {Finethy}, {Fox}, {Garrison}, {Gibbons}, {Goldstein}, {Gommers},
  {Greco}, {Greenfield}, {Groener}, {Grollier}, {Hagen}, {Hirst}, {Homeier},
  {Horton}, {Hosseinzadeh}, {Hu}, {Hunkeler}, {Ivezi{\'c}}, {Jain}, {Jenness},
  {Kanarek}, {Kendrew}, {Kern}, {Kerzendorf}, {Khvalko}, {King}, {Kirkby},
  {Kulkarni}, {Kumar}, {Lee}, {Lenz}, {Littlefair}, {Ma}, {Macleod},
  {Mastropietro}, {McCully}, {Montagnac}, {Morris}, {Mueller}, {Mumford},
  {Muna}, {Murphy}, {Nelson}, {Nguyen}, {Ninan}, {N{\"o}the}, {Ogaz}, {Oh},
  {Parejko}, {Parley}, {Pascual}, {Patil}, {Patil}, {Plunkett}, {Prochaska},
  {Rastogi}, {Reddy Janga}, {Sabater}, {Sakurikar}, {Seifert}, {Sherbert},
  {Sherwood-Taylor}, {Shih}, {Sick}, {Silbiger}, {Singanamalla}, {Singer},
  {Sladen}, {Sooley}, {Sornarajah}, {Streicher}, {Teuben}, {Thomas},
  {Tremblay}, {Turner}, {Terr{\'o}n}, {van Kerkwijk}, {de la Vega}, {Watkins},
  {Weaver}, {Whitmore}, {Woillez}, {Zabalza}, \& {Contributors}}]{astropy:2018}
{Price-Whelan}, A.~M., {Sip{\H{o}}cz}, B.~M., {G{\"u}nther}, H.~M., {et~al.}
  2018, AJ, 156, 123, \dodoi{10.3847/1538-3881/aabc4f}

\bibitem[{{Samus} {et~al.}(2017){Samus}, {Kazarovets}, {Durlevich}, {Kireeva},
  \& {Pastukhova}}]{gcvs}
{Samus}, N.~N., {Kazarovets}, E.~V., {Durlevich}, O.~V., {Kireeva}, N.~N., \&
  {Pastukhova}, E.~N. 2017, Astronomy Reports, 61, 80,
  \dodoi{10.1134/S1063772917010085}

\bibitem[{{Schlafly} \& {Finkbeiner}(2011)}]{s-f-2011}
{Schlafly}, E.~F., \& {Finkbeiner}, D.~P. 2011, ApJ, 737, 103,
  \dodoi{10.1088/0004-637X/737/2/103}

\bibitem[{{Schwarzschild} \& {H{\"a}rm}(1967)}]{schwarzschild1967}
{Schwarzschild}, M., \& {H{\"a}rm}, R. 1967, ApJ, 150, 961,
  \dodoi{10.1086/149396}

\bibitem[{{Smelcer}(2002)}]{smelcer2002}
{Smelcer}, L. 2002, Information Bulletin on Variable Stars, 5323, 1

\bibitem[{{Smelcer}(2006)}]{smelcer2006}
---. 2006, Open European Journal on Variable Stars, 23, 3

\bibitem[{{Smolec}(2016)}]{smolec-2016}
{Smolec}, R. 2016, MNRAS, 456, 3475, \dodoi{10.1093/mnras/stv2868}

\bibitem[{{Smolec} \& {Moskalik}(2008{\natexlab{a}})}]{warsawcode-2008}
{Smolec}, R., \& {Moskalik}, P. 2008{\natexlab{a}}, Acta Astr., 58, 193.
\newblock \doarXiv{0809.1979}

\bibitem[{{Smolec} \& {Moskalik}(2008{\natexlab{b}})}]{smolec-moskalik-2008}
---. 2008{\natexlab{b}}, Acta Astr., 58, 233.
\newblock \doarXiv{0809.1986}

\bibitem[{{Snellen} {et~al.}(2012){Snellen}, {Stuik}, {Navarro}, {Bettonvil},
  {Kenworthy}, {de Mooij}, {Otten}, {ter Horst}, \& {le Poole}}]{mascara-2012}
{Snellen}, I. A.~G., {Stuik}, R., {Navarro}, R., {et~al.} 2012, in Society of
  Photo-Optical Instrumentation Engineers (SPIE) Conference Series, Vol. 8444,
  Ground-based and Airborne Telescopes IV, 84440I

\bibitem[{{Szab{\'o}} {et~al.}(2004){Szab{\'o}}, {Koll{\'a}th}, \&
  {Buchler}}]{szabo-2004}
{Szab{\'o}}, R., {Koll{\'a}th}, Z., \& {Buchler}, J.~R. 2004, \aap, 425, 627,
  \dodoi{10.1051/0004-6361:20035698}

\bibitem[{{Szatm{\'a}ry} {et~al.}(2003){Szatm{\'a}ry}, {Kiss}, \&
  {Bebesi}}]{tumi-szatmary2003}
{Szatm{\'a}ry}, K., {Kiss}, L.~L., \& {Bebesi}, Z. 2003, \aap, 398, 277,
  \dodoi{10.1051/0004-6361:20021646}

\bibitem[{{Tashibu} {et~al.}(2017){Tashibu}, {Yasuda}, \& {Kozasa}}]{Tashibu}
{Tashibu}, S., {Yasuda}, Y., \& {Kozasa}, T. 2017, MNRAS, 466, 1709,
  \dodoi{10.1093/mnras/stw3160}

\bibitem[{{Templeton} {et~al.}(2005){Templeton}, {Mattei}, \&
  {Willson}}]{templeton2005}
{Templeton}, M.~R., {Mattei}, J.~A., \& {Willson}, L.~A. 2005, AJ, 130, 776,
  \dodoi{10.1086/431740}

\bibitem[{{Townsend} \& {Teitler}(2013)}]{GYRE}
{Townsend}, R.~H.~D., \& {Teitler}, S.~A. 2013, MNRAS, 435, 3406,
  \dodoi{10.1093/mnras/stt1533}

\bibitem[{{Trabucchi} {et~al.}(2019){Trabucchi}, {Wood}, {Montalb{\'a}n},
  {Marigo}, {Pastorelli}, \& {Girardi}}]{colibri2019}
{Trabucchi}, M., {Wood}, P.~R., {Montalb{\'a}n}, J., {et~al.} 2019, MNRAS, 482,
  929, \dodoi{10.1093/mnras/sty2745}

\bibitem[{{Uttenthaler}(2013)}]{uttenthaler-massloss}
{Uttenthaler}, S. 2013, \aap, 556, A38, \dodoi{10.1051/0004-6361/201321196}

\bibitem[{{Uttenthaler} {et~al.}(2015){Uttenthaler}, {Blommaert}, {Wood},
  {Lebzelter}, {Aringer}, {Schultheis}, \& {Ryde}}]{uttenthaler-2015}
{Uttenthaler}, S., {Blommaert}, J.~A.~D.~L., {Wood}, P.~R., {et~al.} 2015,
  MNRAS, 451, 1750, \dodoi{10.1093/mnras/stv1052}

\bibitem[{{Uttenthaler} {et~al.}(2016{\natexlab{a}}){Uttenthaler}, {Greimel},
  \& {Templeton}}]{uttenthaler2016}
{Uttenthaler}, S., {Greimel}, R., \& {Templeton}, M. 2016{\natexlab{a}},
  Astronomische Nachrichten, 337, 293, \dodoi{10.1002/asna.201512296}

\bibitem[{{Uttenthaler} {et~al.}(2019){Uttenthaler}, {McDonald}, {Bernhard},
  {Cristallo}, \& {Gobrecht}}]{uttenthaler-massloss-2}
{Uttenthaler}, S., {McDonald}, I., {Bernhard}, K., {Cristallo}, S., \&
  {Gobrecht}, D. 2019, \aap, 622, A120, \dodoi{10.1051/0004-6361/201833794}

\bibitem[{{Uttenthaler} {et~al.}(2016{\natexlab{b}}){Uttenthaler}, {Meingast},
  {Lebzelter}, {Aringer}, {Joyce}, {Hinkle}, {Guzman-Ramirez}, \&
  {Greimel}}]{lxcyg-2016}
{Uttenthaler}, S., {Meingast}, S., {Lebzelter}, T., {et~al.}
  2016{\natexlab{b}}, \aap, 585, A145, \dodoi{10.1051/0004-6361/201526619}

\bibitem[{{Uttenthaler} {et~al.}(2011){Uttenthaler}, {van Stiphout}, {Voet},
  {van Winckel}, {van Eck}, {Jorissen}, {Kerschbaum}, {Raskin}, {Prins},
  {Pessemier}, {Waelkens}, {Fr{\'e}mat}, {Hensberge}, {Dumortier}, \&
  {Lehmann}}]{tumi-uttenthaler2011}
{Uttenthaler}, S., {van Stiphout}, K., {Voet}, K., {et~al.} 2011, \aap, 531,
  A88, \dodoi{10.1051/0004-6361/201116463}

\bibitem[{{van der Walt} {et~al.}(2011){van der Walt}, {Colbert}, \&
  {Varoquaux}}]{numpy}
{van der Walt}, S., {Colbert}, S.~C., \& {Varoquaux}, G. 2011, Computing in
  Science Engineering, 13, 22, \dodoi{10.1109/MCSE.2011.37}

\bibitem[{{Ventura} {et~al.}(2018){Ventura}, {Karakas}, {Dell'Agli},
  {Garc{\'{\i}}a-Hern{\'a}ndez}, \& {Guzman-Ramirez}}]{VenturaKarakas}
{Ventura}, P., {Karakas}, A., {Dell'Agli}, F., {Garc{\'{\i}}a-Hern{\'a}ndez},
  D.~A., \& {Guzman-Ramirez}, L. 2018, MNRAS, 475, 2282,
  \dodoi{10.1093/mnras/stx3338}

\bibitem[{{Whitelock}(2012)}]{whitelock-2012}
{Whitelock}, P.~A. 2012, Astroph. Space Sci., 341, 123,
  \dodoi{10.1007/s10509-012-0993-x}

\bibitem[{{Whitelock} {et~al.}(2008){Whitelock}, {Feast}, \& {Van
  Leeuwen}}]{whitelock2008}
{Whitelock}, P.~A., {Feast}, M.~W., \& {Van Leeuwen}, F. 2008, MNRAS, 386, 313,
  \dodoi{10.1111/j.1365-2966.2008.13032.x}

\bibitem[{{Wood}(2007)}]{wood-2007}
{Wood}, P.~R. 2007, in Astronomical Society of the Pacific Conference Series,
  Vol. 362, The Seventh Pacific Rim Conference on Stellar Astrophysics, ed.
  Y.~W. {Kang}, H.~W. {Lee}, K.~C. {Leung}, \& K.~S. {Cheng}, 234

\bibitem[{{Wood} \& {Zarro}(1981)}]{wood-zarro-1981}
{Wood}, P.~R., \& {Zarro}, D.~M. 1981, ApJ, 247, 247, \dodoi{10.1086/159032}

\bibitem[{{Ya'Ari} \& {Tuchman}(1996)}]{yaari1996}
{Ya'Ari}, A., \& {Tuchman}, Y. 1996, ApJ, 456, 350, \dodoi{10.1086/176656}

\end{thebibliography}
%%%%%%%%%%%%%%%%%%%%%%%%%%%%%%%%%%%%%%%%%%%%%%%%%%

%%%%%%%%%%%%%%%%% APPENDICES %%%%%%%%%%%%%%%%%%%%%

\appendix

\begin{figure}
\centering
	\includegraphics[width=0.9\textwidth]{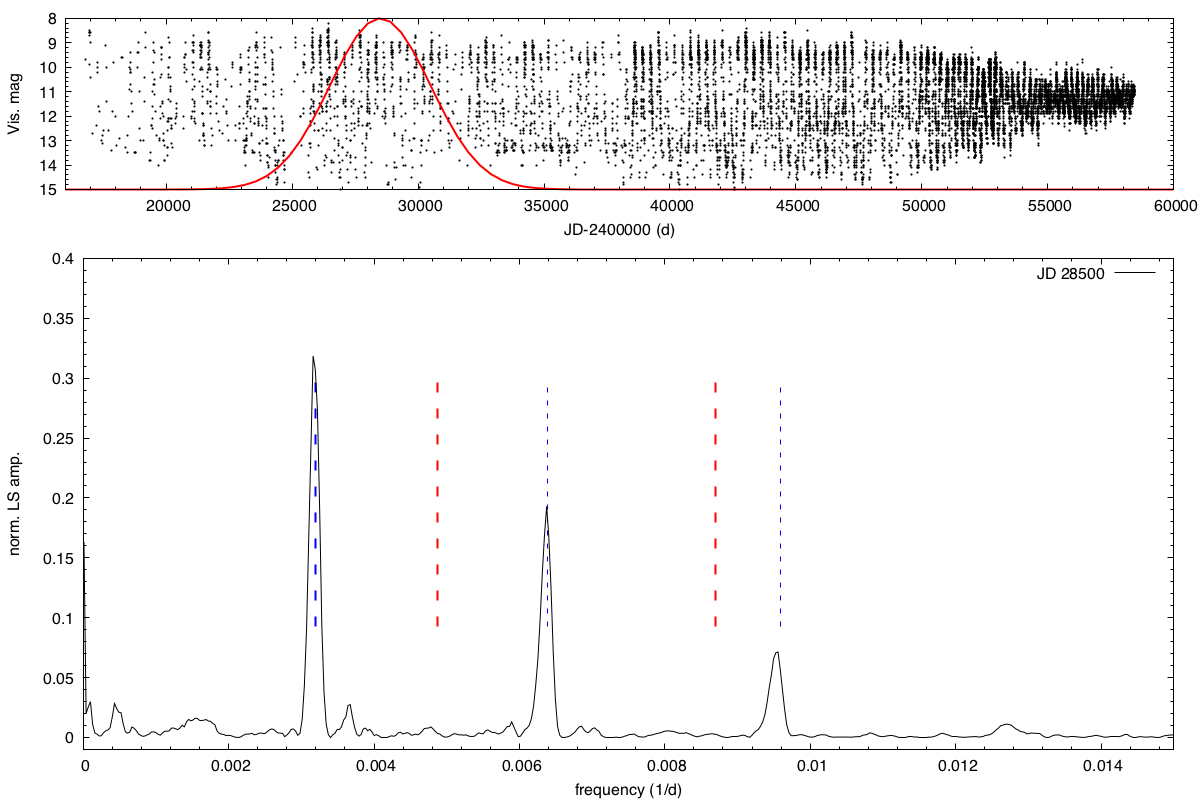}
    \caption{Still frame from the animation showing the evolution of the frequency content of the star. Top: the AAVSO light curve; the overlaid red line shows the position and width of the filter. Bottom: the corresponding LS frequency spectrum. Blue lines mark the original pulsation frequency and its two harmonics. Red marks the current period of the fundamental mode and the position of the first overtone.}
    \label{anim-still}
\end{figure}

Figure \ref{anim-still} shows the frequency spectrum of one segment of the light curve weighted with a Gaussian filter. The animated version of the evolution of successive frequency spectra calculated with a sliding Gaussian window to the data is available in the online version of the journal and can also be accessed here: \url{https://www.youtube.com/watch?v=115DQJM_KBA}. The width of the filter is $\sigma = 2000$~d, the time step is 200~d.

\begin{figure*}
	\includegraphics[width=\textwidth]{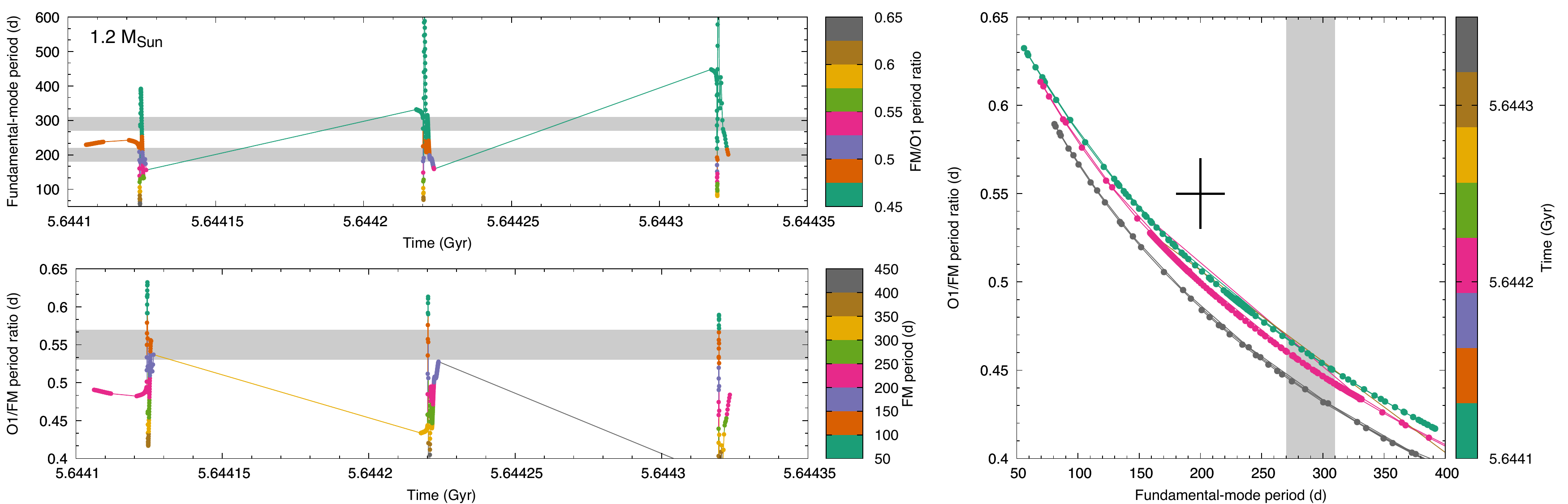}
    \caption{TP-AGB diagnostic diagram of the 1.2 $\mathrm{M}_\odot$ model. }
    \label{diag-1.2}
\end{figure*}

\begin{figure*}
	\includegraphics[width=\textwidth]{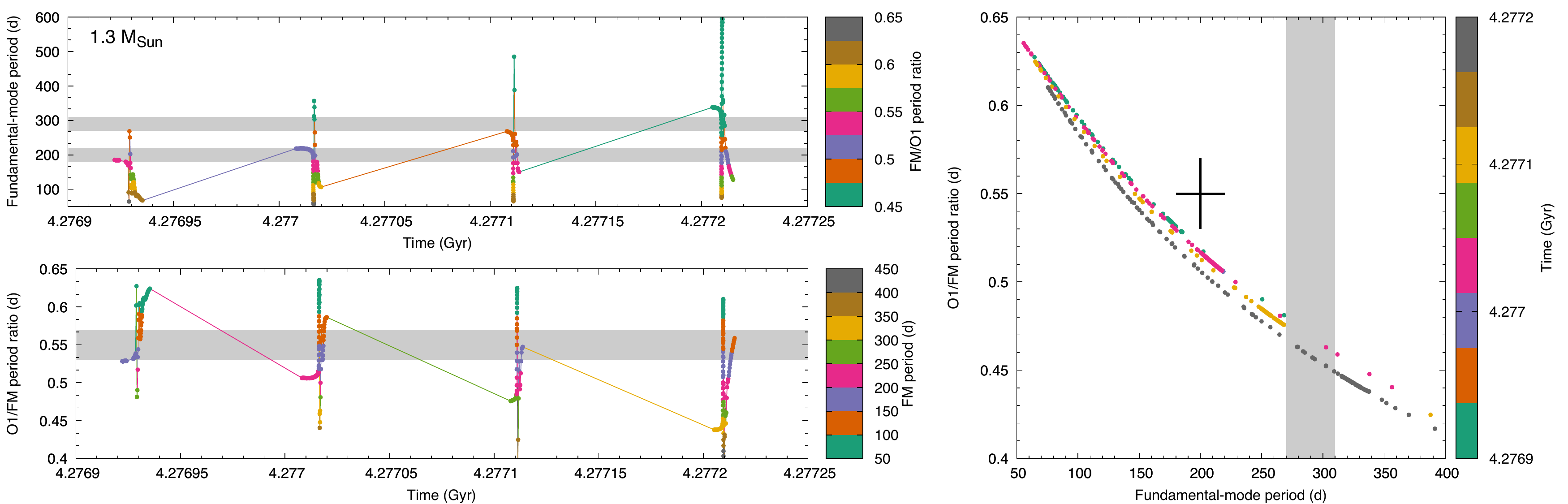}
    \caption{TP-AGB diagnostic diagram of the 1.3 $\mathrm{M}_\odot$ model. }
    \label{diag-1.3}
\end{figure*}

\begin{figure*}
	\includegraphics[width=\textwidth]{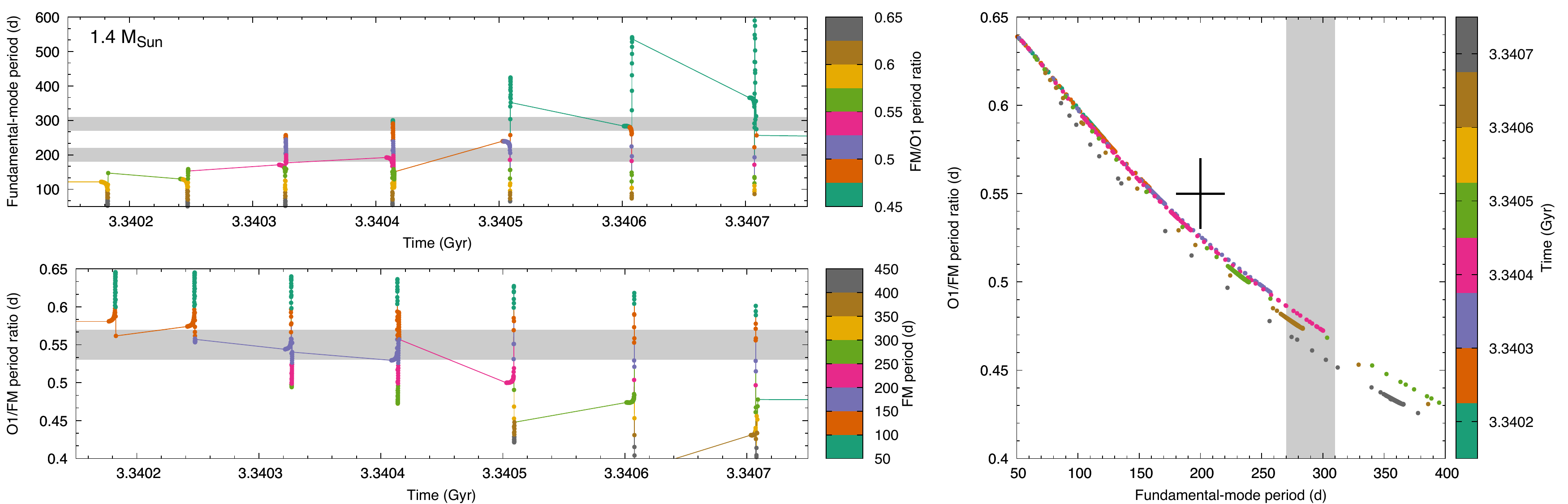}
    \caption{TP-AGB diagnostic diagram of the 1.4 $\mathrm{M}_\odot$ model. }
    \label{diag-1.4}
\end{figure*}

\begin{figure*}
	\includegraphics[width=\textwidth]{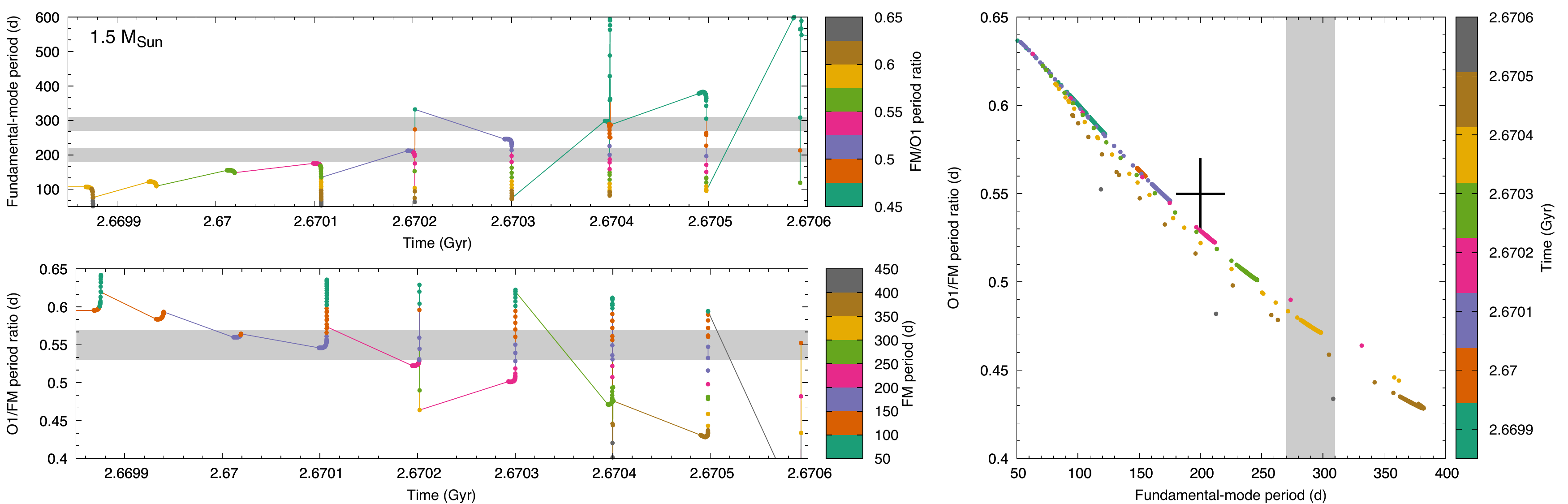}
    \caption{TP-AGB diagnostic diagram of the 1.5 $\mathrm{M}_\odot$ model. }
    \label{diag-1.5}
\end{figure*}

\begin{figure*}
	\includegraphics[width=\textwidth]{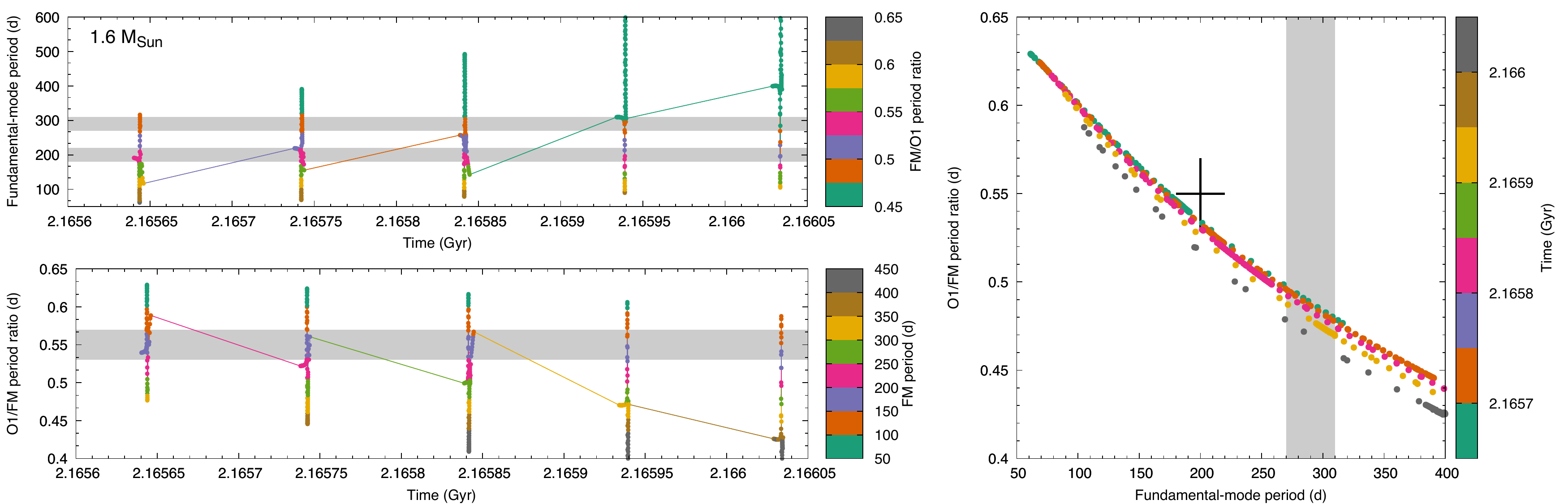}
    \caption{TP-AGB diagnostic diagram of the 1.6 $\mathrm{M}_\odot$ model. }
    \label{diag-1.6}
\end{figure*}

\begin{figure*}
	\includegraphics[width=\textwidth]{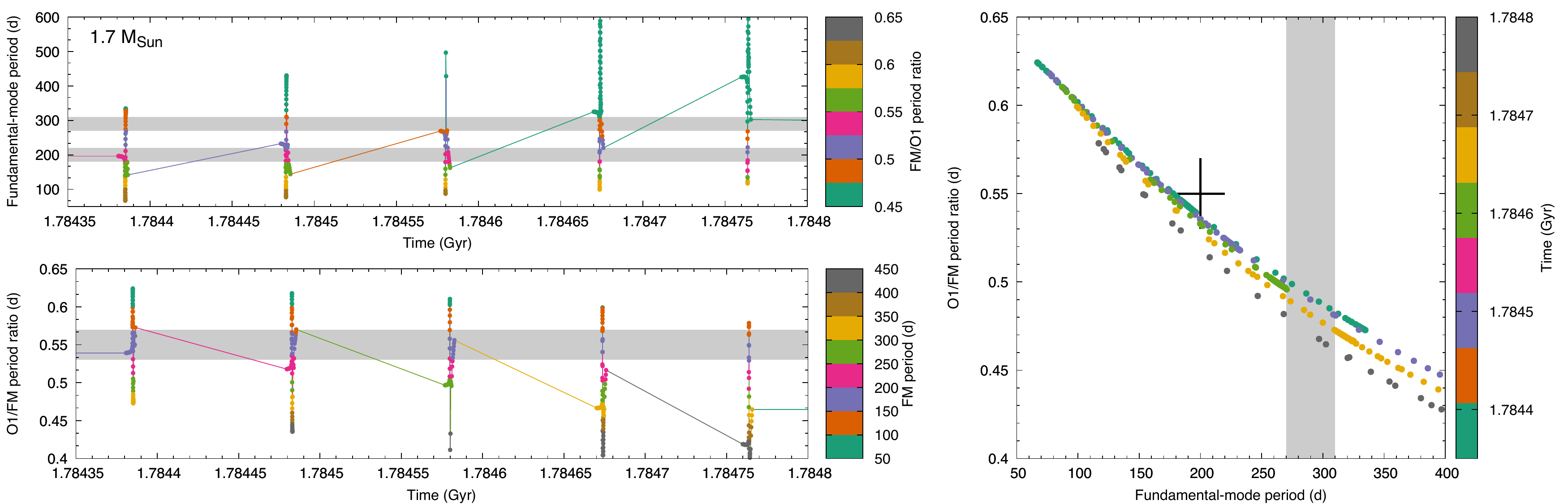}
    \caption{TP-AGB diagnostic diagram of the 1.7 $\mathrm{M}_\odot$ model. }
    \label{diag-1.7}
\end{figure*}

\begin{figure*}
	\includegraphics[width=\textwidth]{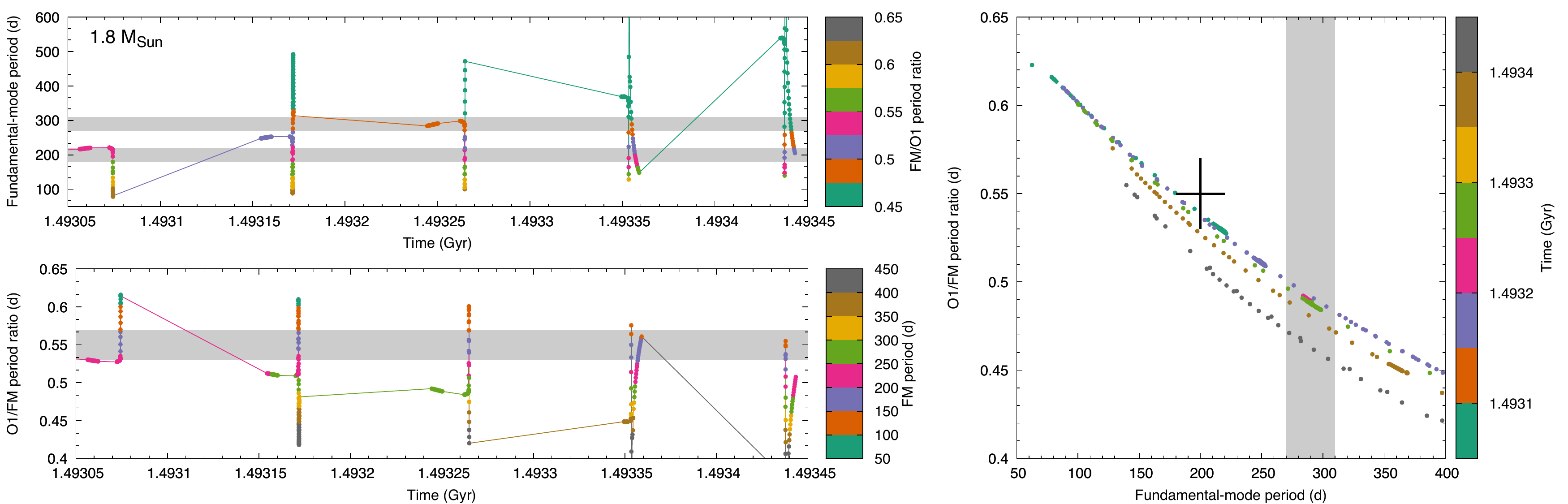}
    \caption{TP-AGB diagnostic diagram of the 1.8 $\mathrm{M}_\odot$ model. }
    \label{diag-1.8}
\end{figure*}

\begin{figure*}
	\includegraphics[width=\textwidth]{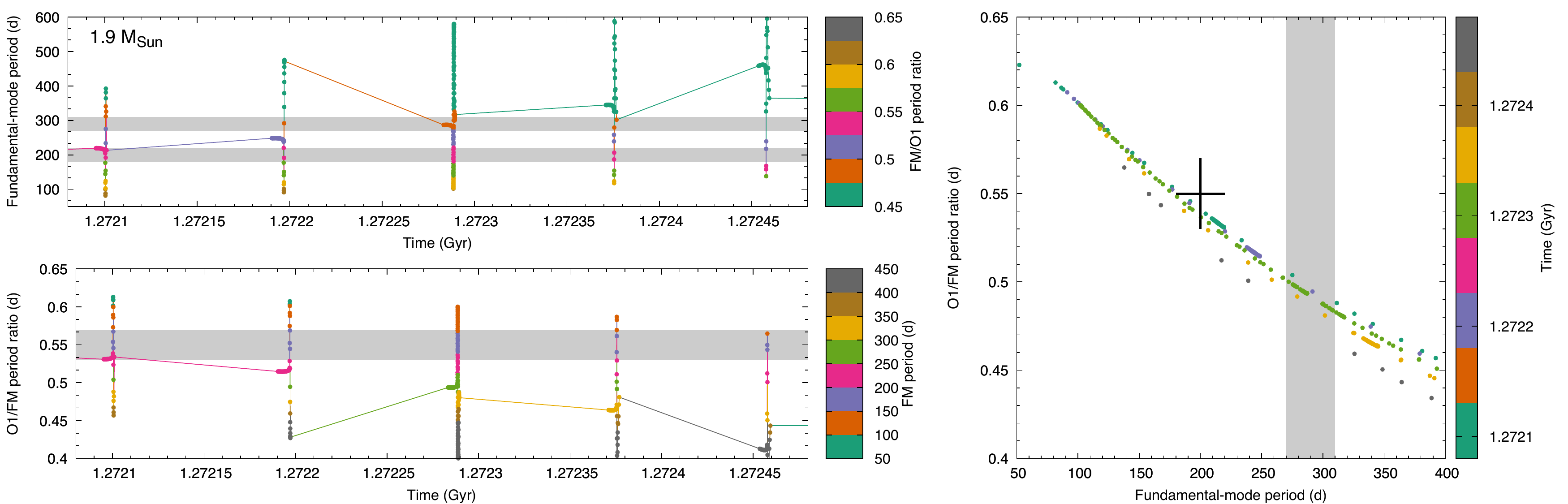}
    \caption{TP-AGB diagnostic diagram of the 1.9 $\mathrm{M}_\odot$ model. }
    \label{diag-1.9}
\end{figure*}

\begin{figure*}
	\includegraphics[width=\textwidth]{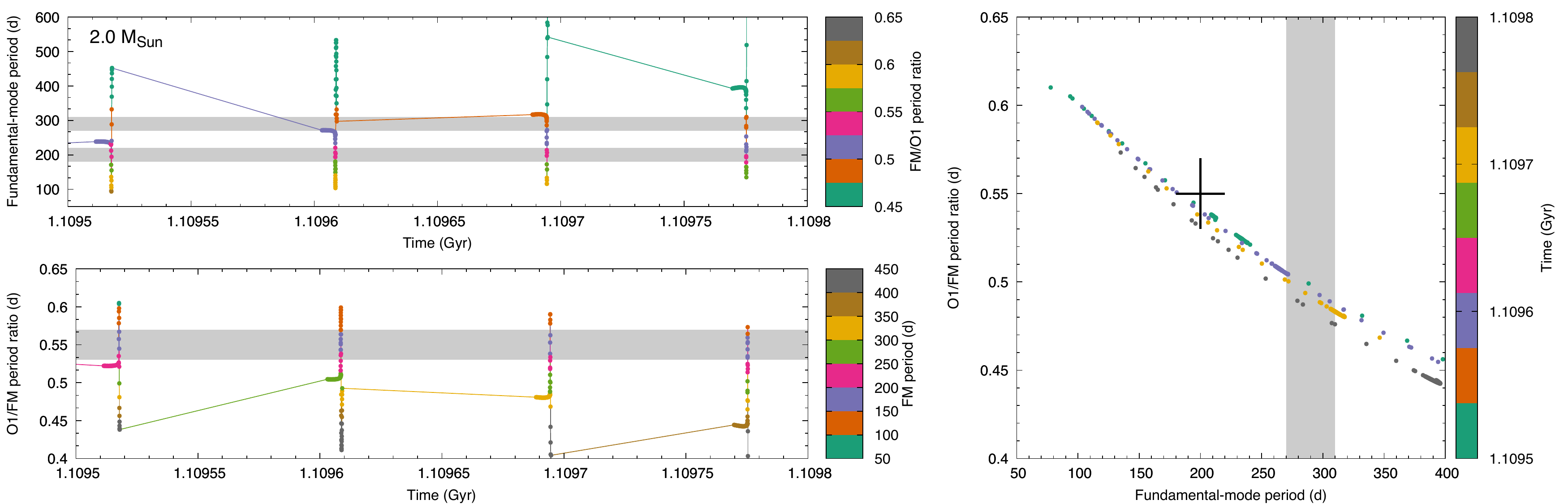}
    \caption{TP-AGB diagnostic diagram of the 2.0 $\mathrm{M}_\odot$ model. }
    \label{diag-2.0}
\end{figure*}

\begin{figure*}
	\includegraphics[width=\textwidth]{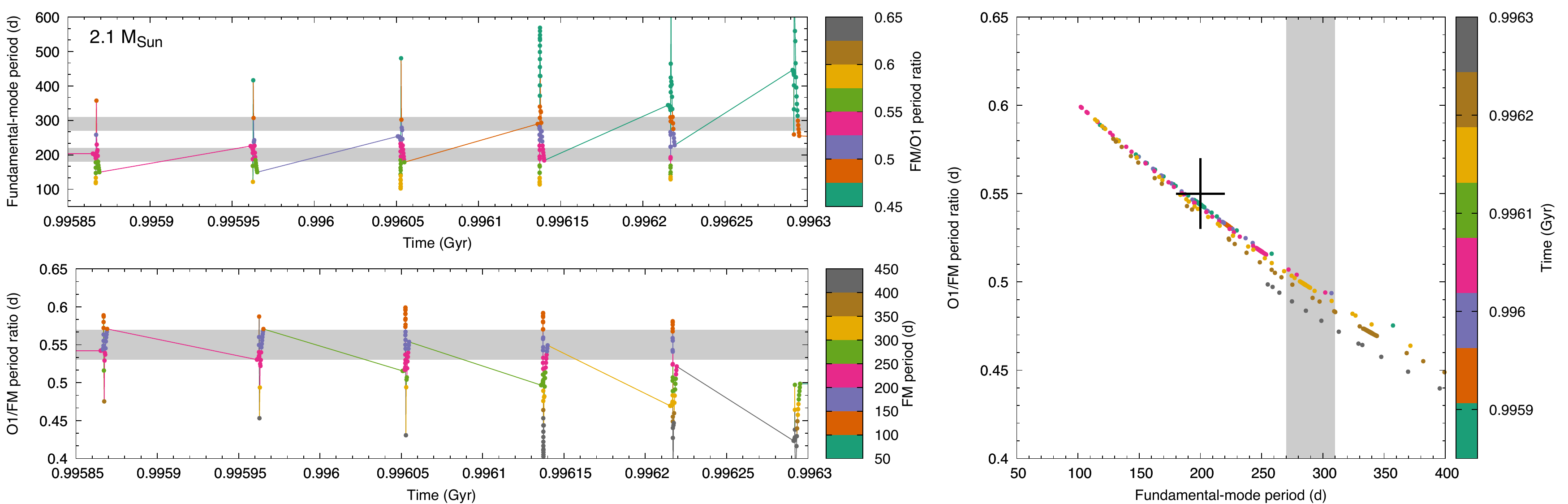}
    \caption{TP-AGB diagnostic diagram of the 2.1 $\mathrm{M}_\odot$ model. }
    \label{diag-2.1}
\end{figure*}

\begin{figure*}
	\includegraphics[width=\textwidth]{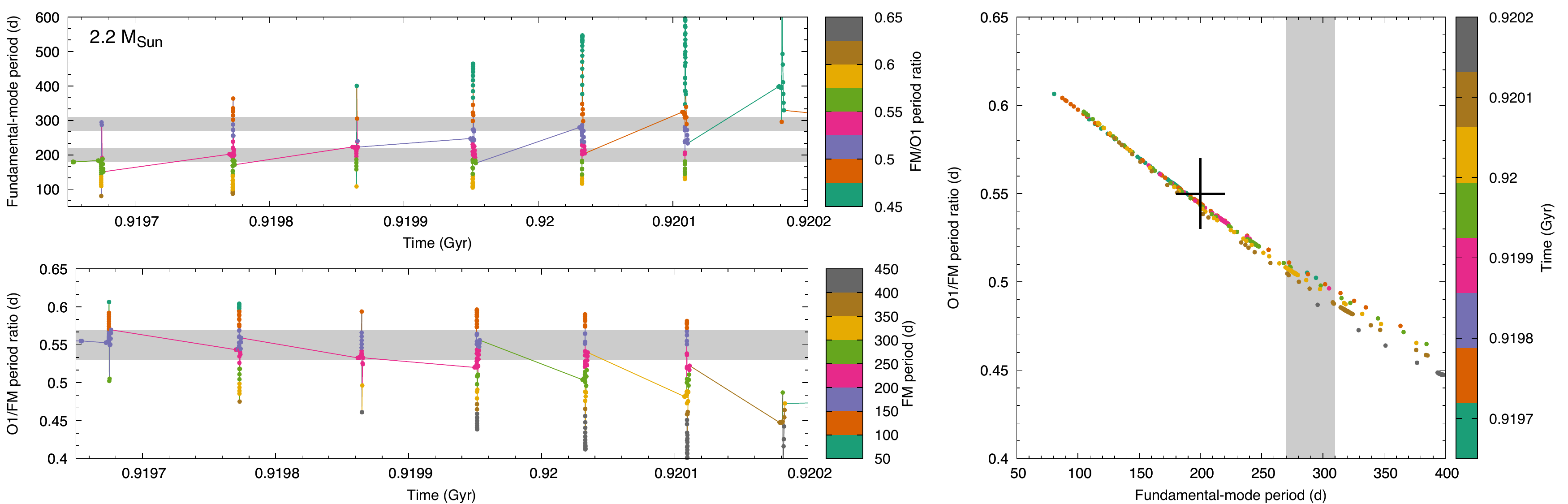}
    \caption{TP-AGB diagnostic diagram of the 2.2 $\mathrm{M}_\odot$ model. }
    \label{diag-2.2}
\end{figure*}

\begin{figure*}
	\includegraphics[width=\textwidth]{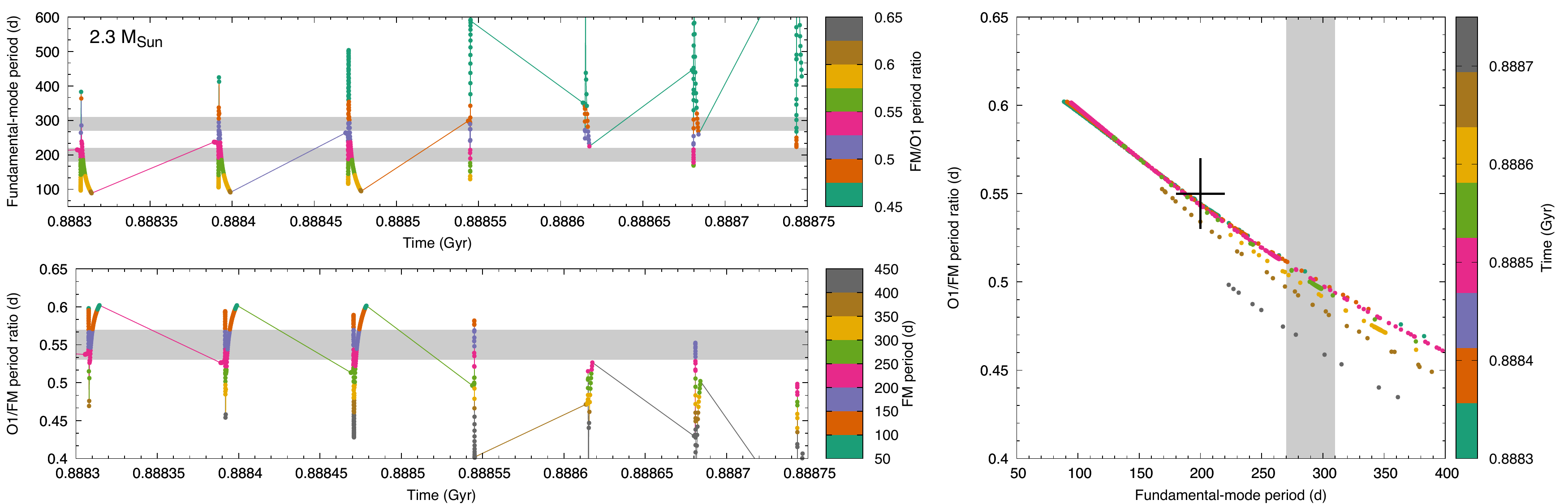}
    \caption{TP-AGB diagnostic diagram of the 2.3 $\mathrm{M}_\odot$ model. }
    \label{diag-2.3}
\end{figure*}

\begin{figure*}
	\includegraphics[width=\textwidth]{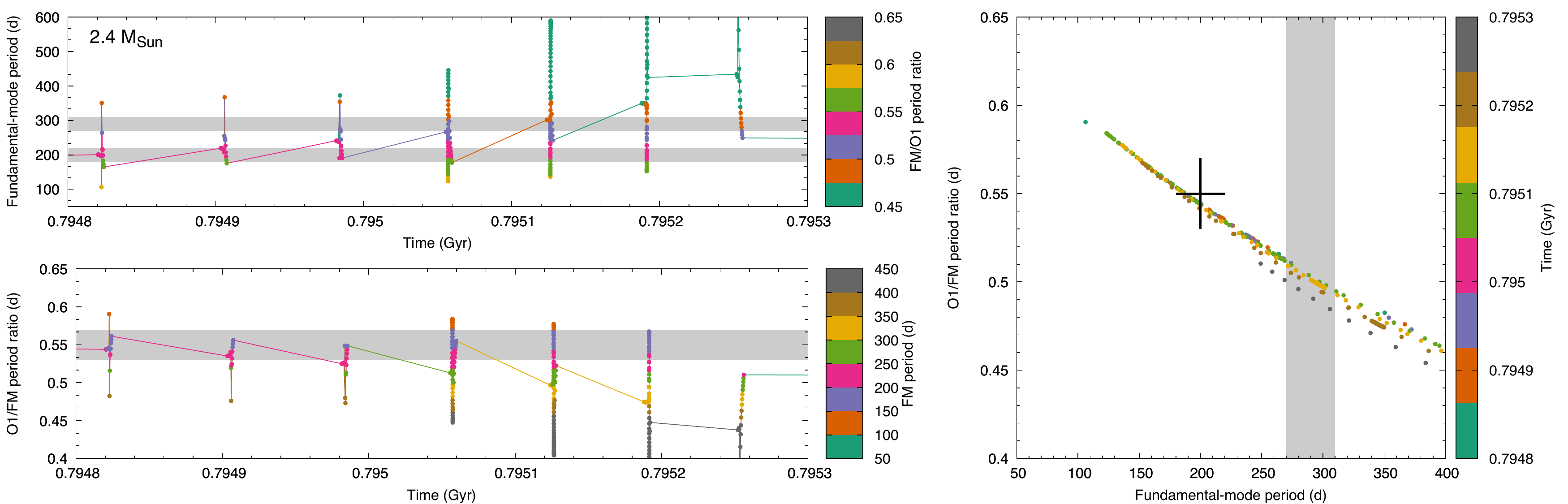}
    \caption{TP-AGB diagnostic diagram of the 2.4 $\mathrm{M}_\odot$ model. }
    \label{diag-2.4}
\end{figure*}

\begin{figure*}
	\includegraphics[width=\textwidth]{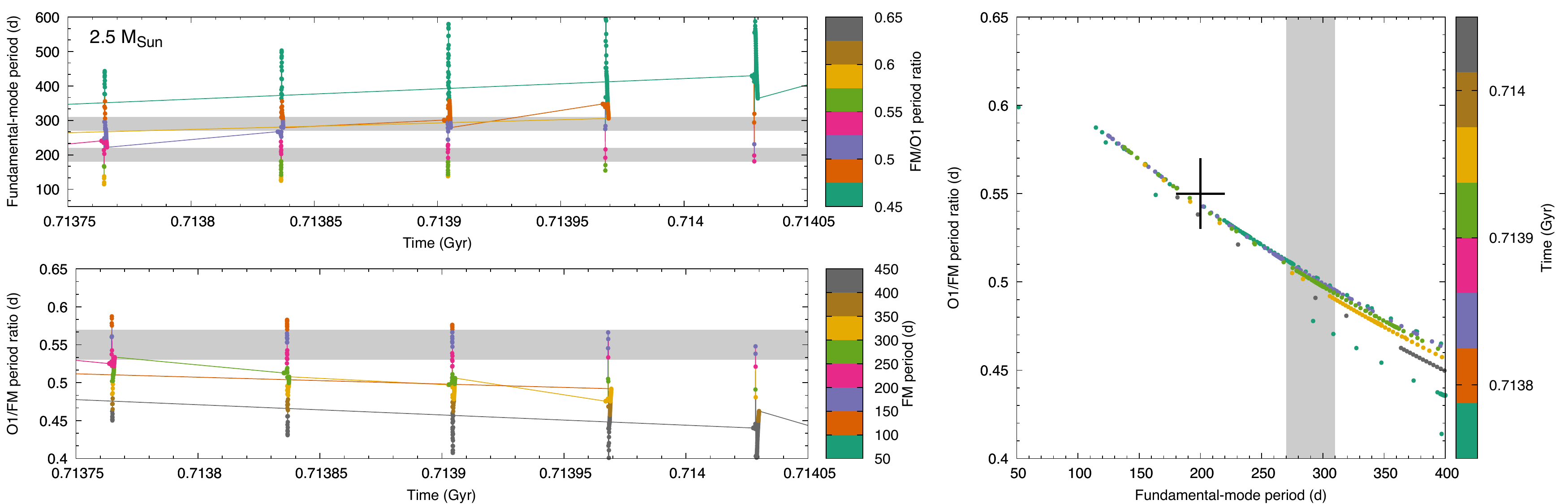}
    \caption{TP-AGB diagnostic diagram of the 2.5 $\mathrm{M}_\odot$ model. }
    \label{diag-2.5}
\end{figure*}

\begin{figure*}
	\includegraphics[width=\textwidth]{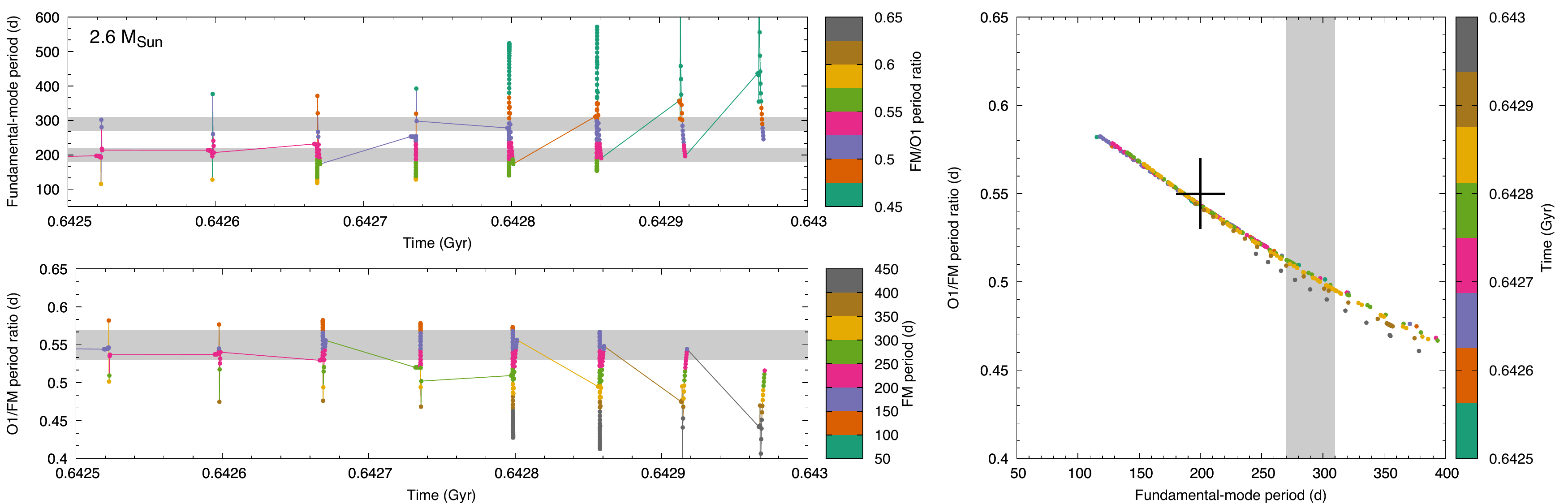}
    \caption{TP-AGB diagnostic diagram of the 2.6 $\mathrm{M}_\odot$ model. }
    \label{diag-2.6}
\end{figure*}

\begin{figure*}
	\includegraphics[width=\textwidth]{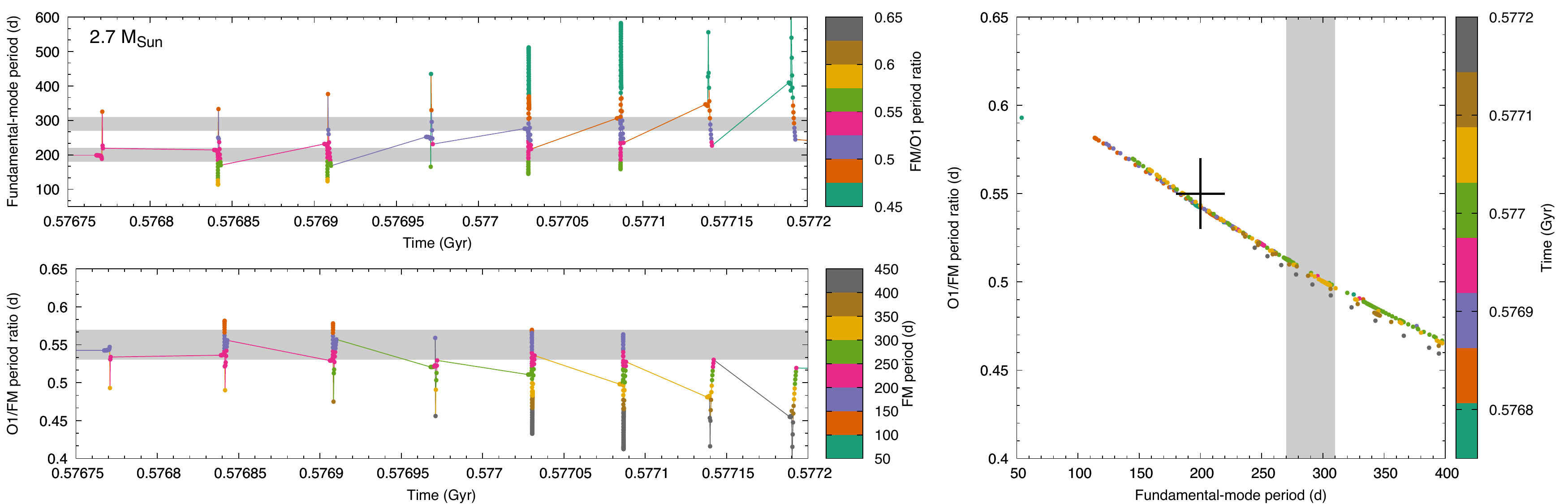}
    \caption{TP-AGB diagnostic diagram of the 2.7 $\mathrm{M}_\odot$ model. }
    \label{diag-2.7}
\end{figure*}

\begin{figure*}
	\includegraphics[width=\textwidth]{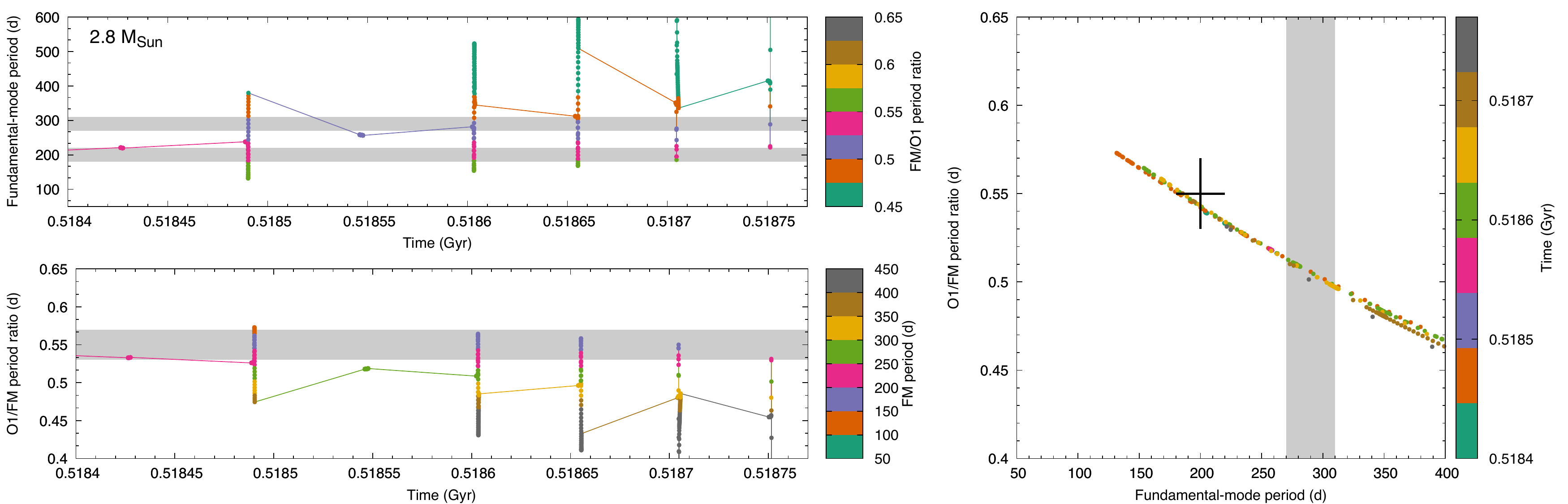}
    \caption{TP-AGB diagnostic diagram of the 2.8 $\mathrm{M}_\odot$ model. }
    \label{diag-2.8}
\end{figure*}

\begin{figure*}
	\includegraphics[width=\textwidth]{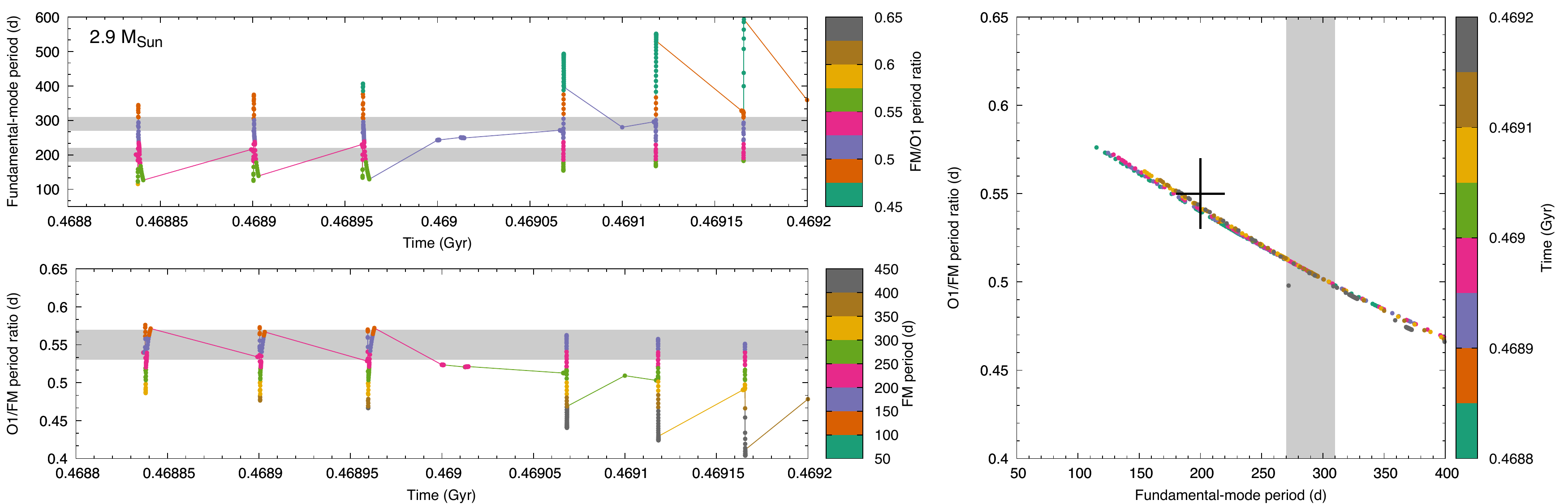}
    \caption{TP-AGB diagnostic diagram of the 2.9 $\mathrm{M}_\odot$ model. }
    \label{diag-2.9}
\end{figure*}

\begin{figure*}
	\includegraphics[width=\textwidth]{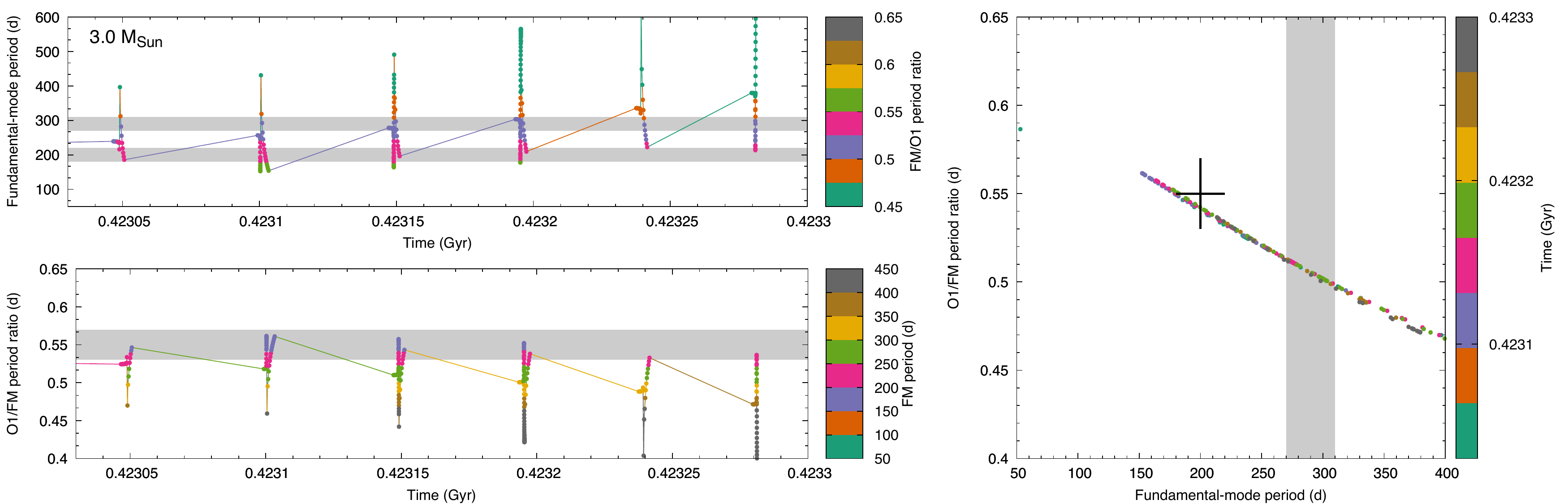}
    \caption{TP-AGB diagnostic diagram of the 3.0 $\mathrm{M}_\odot$ model. }
    \label{diag-3.0}
\end{figure*}

We created diagnostic plots that show the evolution of the fundamental period and the O1/FM period ratio against time, and against each other for each mass step (Figs.~\ref{diag-1.2} to \ref{diag-3.0}).

\label{lastpage}
\end{document}